\documentclass[twocolumn,showpacs,amsmath,amsfonts,amssymb,superscriptaddress,nofootinbib,preprintnumbers]{revtex4-1}

\pdfoutput=1
\usepackage[usenames,dvipsnames]{color}
\usepackage[colorlinks=true, linkcolor=BrickRed, citecolor=Blue, urlcolor=Blue, filecolor=Blue]{hyperref}
\usepackage{longtable,epsfig,graphicx,verbatim,xspace,multirow,ulem}

\newcommand{\nudm}{$\nu$CDM\xspace}
\newcommand{\lcdm}{$\Lambda$CDM\xspace}
\newcommand{\Planck}{{\it Planck}\xspace}
\newcommand{\WiggleZ}{{\it WiggleZ}\xspace}

\begin{document}

\title{Exploring dark matter microphysics with galaxy surveys}

\author{Miguel Escudero}
\affiliation{Instituto de F\'{\i}sica Corpuscular (IFIC)$,$
CSIC-Universitat de Val\`encia$,$   
Apartado de Correos 22085$,$ E-46071 Valencia$,$ Spain}
\author{Olga Mena}
\affiliation{Instituto de F\'{\i}sica Corpuscular (IFIC)$,$
CSIC-Universitat de Val\`encia$,$
Apartado de Correos 22085$,$ E-46071 Valencia$,$ Spain}
\author{Aaron C. Vincent}
\affiliation{Institute for Particle Physics Phenomenology (IPPP), Durham University, Durham DH1 3LE, UK}
\author{Ryan J. Wilkinson}
\affiliation{Institute for Particle Physics Phenomenology (IPPP), Durham University, Durham DH1 3LE, UK}
\author{C\'eline B\oe hm}
\affiliation{Institute for Particle Physics Phenomenology (IPPP), Durham University, Durham DH1 3LE, UK}


\begin{abstract}
We use present cosmological observations and forecasts of future experiments to illustrate the power of large-scale structure (LSS) surveys in probing dark matter (DM) microphysics and unveiling potential deviations from the standard \lcdm scenario. To quantify this statement, we focus on an extension of \lcdm with DM--neutrino scattering, which leaves a distinctive imprint on the angular and matter power spectra. After finding that future CMB experiments (such as COrE+) will not significantly improve the constraints set by the \Planck satellite, we show that the next generation of galaxy clustering surveys (such as DESI) could play a leading role in constraining alternative cosmologies and even have the potential to make a discovery. Typically we find that DESI would be an order of magnitude more sensitive to DM interactions than \Planck, thus probing effects that until now have only been accessible via $N$-body simulations.
\end{abstract}
\preprint{IFIC/15-32, IPPP/15/29, DCPT/15/58}

\date{\today}

\maketitle

\section{Introduction}
\label{sec:intro}

Dark matter (DM) is required to explain the galactic rotation curves, lensing and virial motions of galaxy clusters, observed matter power spectrum and cosmic microwave background (CMB) acoustic peaks. The current paradigm is that DM can be well-approximated by a collisionless fluid, consisting of weakly-interacting massive particles (WIMPs) and leading to a characteristic matter power spectrum (${\rm P}(k) \propto k^{-3}$). However, direct evidence for WIMPs remains elusive and it is now legitimate to question the validity of the standard picture.

The \lcdm framework provides an excellent fit to the currently-available data~\cite{Ade:2013zuv,Adam:2015rua}. Any future rejection or modification of this model must therefore come in the form of comparison with a (better) alternative hypothesis, or through the discovery of new data in direct conflict with \lcdm predictions. Possible modifications to the Standard Model either assume new relativistic species such as sterile neutrinos~\cite{Dodelson:1993je,Lesgourgues:2006nd}, or different DM characteristics such as interactions with particles in the visible or dark sector. Either way, these ``Beyond the Standard Model" scenarios predict a modification of the \lcdm matter power spectrum at small scales.

There is in fact a plethora of DM models in the literature that exhibit such deviations~(e.g. Refs.~\cite{Boehm:2000gq,Boehm:2001hm,Boehm:2004th,Bertschinger:2006nq,Mangano:2006mp,Serra:2009uu,
Wilkinson:2014ksa,Aarssen:2012fx,Farzan:2014gza,Boehm:2014vja,Cherry:2014xra,Bertoni:2014mva,Schewtschenko:2014fca,
Davis:2015rza,Sigurdson:2004zp,Wilkinson:2013kia,Dolgov:2013una,Chen:2002yh,Dvorkin:2013cea,Park:2012ru,
Diamanti:2012tg,Blennow:2012de,Cyr-Racine:2013fsa}), many of which predict additional damping and/or oscillations in the ${\rm P}(k)$. While CMB experiments such as \Planck allow one to constrain the cosmological parameters with unprecedented precision~\cite{Ade:2013zuv,Adam:2015rua}, extracting the ${\rm P}(k)$ from \Planck or the next-generation of CMB probes (such as COrE+~\cite{Bouchet:2011ck} or PIXIE~\cite{Kogut:2011xw}) will be limited by the large uncertainties involved in foreground modelling, which hinder any angular power spectrum analysis at large $\ell$\footnote{An additional difficulty is that the $C_\ell$ are the result of the convolution of the ${\rm P}(k)$ with a window (Bessel) function that accounts for the angular scale, thus preventing one from detecting small features in the ${\rm P}(k)$.}.  Therefore, to unravel the nature of DM, a direct probe of the ${\rm P}(k)$ is needed. Here we show that the next generation of large-scale structure (LSS) surveys could provide us with key information on the particle properties of DM, due to their extremely high precision.

Galaxy clustering surveys~\cite{Anderson:2013zyy,Anderson21122012,Beutler01102011,Blake11122011, Padmanabhan11122012,Percival01022010,Parkinson:2012vd} have already observed the imprint of Baryon Acoustic Oscillations (BAOs), a standard ruler to measure the Hubble expansion rate, $H(z)$, and the angular diameter distance, $D_{\rm A}(z)$. Recently, the Baryon Oscillation Spectroscopic Survey (BOSS) collaboration~\cite{Dawson:2012va} reported a separate extraction of $H(z)$ and $D_{\rm A}(z)$ to a precision of 1\%~\cite{Anderson:2013zyy}. Here we show that by exploiting all of the information contained in the shape of the full ${\rm P}(k)$ (rather than solely the BAO geometrical signature~\cite{Hamann:2010pw,Giusarma:2012ph,Giusarma:2013pmn}), one can test the validity of the \lcdm model at scales below a Mpc.

For concreteness, we focus on one specific scenario, in which DM scatters elastically with the active neutrinos (hereafter, \nudm)~\cite{Boehm:2000gq,Boehm:2001hm,Boehm:2004th,Bertschinger:2006nq,Mangano:2006mp,Serra:2009uu,
Wilkinson:2014ksa,Aarssen:2012fx,Farzan:2014gza,Boehm:2014vja,Cherry:2014xra,Bertoni:2014mva,Schewtschenko:2014fca,
Davis:2015rza}. Such a DM candidate erases small-scale perturbations through collisional damping~\cite{Boehm:2000gq,Boehm:2004th} and suppresses neutrino free-streaming in the early universe. This leaves a unique signature in the angular and matter power spectra and provides us with a framework to quantify the potential of future LSS surveys to constrain DM microphysics. We exploit both the current publicly available galaxy power spectrum data (in particular, from the \WiggleZ survey~\cite{Parkinson:2012vd}) and the expected full-shape power spectrum measurements from the forthcoming Dark Energy Spectroscopic Instrument (DESI)~\cite{Levi:2013gra}.

The structure of the paper is as follows. In Sec.~\ref{sec:dmnu}, we describe the \nudm scenario. In Sec.~\ref{sec:currentconstraints}, we compute up-to-date constraints using both CMB data from \Planck and full-shape LSS data from the \WiggleZ survey. In Sec.~\ref{sec:forecasts}, we perform a forecast of the sensitivity of planned experiments such as COrE+ and DESI to the \nudm framework (and any model that generates small deviations from \lcdm). Finally, we draw our main conclusions in Sec.~\ref{sec:conc}.

\section{Dark matter--neutrino interactions}
\label{sec:dmnu}

In the \nudm scenario, DM remains in kinetic contact with the neutrino sector long after the chemical freeze-out (see Refs.~\cite{Boehm:2000gq,Boehm:2001hm,Boehm:2004th,Bertschinger:2006nq,Mangano:2006mp,Serra:2009uu,
Wilkinson:2014ksa,Aarssen:2012fx,Farzan:2014gza,Boehm:2014vja,Cherry:2014xra,Bertoni:2014mva,Schewtschenko:2014fca,
Davis:2015rza} for previous related work). Small-scale DM perturbations are then erased as a result of ongoing elastic scattering through ``collisional damping''~\cite{Boehm:2000gq,Boehm:2004th}, rather than slowly clustering under gravity. At the same time, neutrinos cannot free-stream as efficiently as in \lcdm and behave more like a relativistic perfect fluid. The main consequences are: (i) an enhancement of the CMB acoustic peaks and (ii) a reduction of small-scale power in the matter power spectrum~\cite{Mangano:2006mp,Serra:2009uu,Wilkinson:2014ksa}.

DM--neutrino interactions do not affect the background equations but they do modify the evolution of the DM and neutrino fluctuations. They can be implemented by a modification of the Euler equations. In the conformal Newtonian gauge\footnote{In analogy to the perturbation equations governing baryon--photon interactions, see e.g. Ref.~\cite{Ma:1995ey}.},
\begin{eqnarray}
\label{Eq:nudmequations}
\dot \theta_{\rm DM} &=& \ k^2 \psi \ - \  {\cal H}\theta_{\rm DM} \ - \ S^{-1}\dot \mu (\theta_{\rm DM}  \ - \ \theta_\nu)~  \\
\dot \theta_\nu &=& \ k^2 \psi \ + \ k^2\left(\frac{1}{4}\delta_\nu  \ - \  \sigma_\nu \right) \ - \ \dot \mu (\theta_\nu \ - \ \theta_{\rm DM})  \nonumber \\
\dot \sigma_{\nu} &=& \frac{4}{15} \theta_\nu \ - \ \frac{3}{10}k F_{\nu 3}  \ - \ \frac{9}{10} \dot \mu \sigma_{\nu} \nonumber   \\
\dot F_{\nu \ell} &=& \frac{k}{2\ell+1}\left[ \ell F_{\nu (\ell-1)} - (\ell+1)F_{\nu (\ell+1)}  \right] - \dot \mu F_{\nu \ell}, \; \ell \geq 3 \nonumber 
\end{eqnarray}
where $\delta$, $\theta$ and $\sigma$ are the density, velocity and shear perturbations respectively, $F_{\nu \ell}$ refer to higher ($\ell>2$) neutrino moments, $\psi$ is the gravitational potential, ${\cal H}$ is the conformal Hubble parameter and $S \equiv (3/4) \hspace{0.4ex} \rho_{\rm DM}/\rho_\nu$. 

The key quantity in Eq.~\eqref{Eq:nudmequations} is $\dot \mu \equiv a \hspace{0.4ex} \sigma_{{\rm DM}-\nu} \hspace{0.4ex} n_{\rm DM}$, which can be written in terms of the dimensionless quantity $u$ defined as
\begin{equation}
u \equiv \left[\frac{\sigma_{\rm{DM}-\nu}}{\sigma_{\rm Th}} \right] \left[\frac{m_{\rm{DM}}}{100~\rm{GeV}} \right]^{- 1}~.
\label{eq:u_nudm}
\end{equation}
This variable $u$ describes the ratio of the DM--neutrino elastic scattering cross section, $\sigma_{{\rm DM}-\nu}$, to the DM mass, $m_{\rm DM}$, normalised to the Thomson cross section, $\sigma_{\rm Th}$ (see Ref.~\cite{Boehm:2001hm}). In our analyses, we will consider both s-wave ($\sigma_{{\rm DM}-\nu,0} =$ constant) and p-wave ($\sigma_{{\rm DM}-\nu,2} \propto T^2$) cross sections. For p-wave cross sections, we can write $u(a) = u \hspace{0.4ex} a^{-2}$, where $u$ is the present-day value and $a$ is the cosmological scale factor, normalised to unity today. The larger the value of $u$, the greater the suppression in the linear matter power spectrum with respect to \lcdm, for a given wavenumber $k$, as shown in Fig.~\ref{Fig:WiggleZ} (and can also be seen in Refs.~\cite{Mangano:2006mp,Serra:2009uu,Wilkinson:2014ksa}).

DM interactions leave a further imprint in the galaxy power spectrum through damped acoustic oscillations, which, in general, show up at smaller scales than those illustrated in Fig.~\ref{Fig:WiggleZ}. They were first pointed out in the context of DM--photon interactions (in the weak-coupling regime) in Ref.~\cite{Boehm:2001hm} and were later observed in interactions with baryons~\cite{Chen:2002yh}, neutrinos~\cite{Mangano:2006mp,Serra:2009uu} and dark radiation~\cite{Diamanti:2012tg,Blennow:2012de}. They arise because the DM fluid acquires a non-zero pressure as a result of interactions with the thermal bath and are therefore similar to the photon--baryon fluid before recombination. Although they cannot be observed using current data, they provide a characteristic signature for future experiments.

However, in addition to the models mentioned above, damped oscillations in the ${\rm P}(k)$ are also expected for certain types of self-interacting DM~\cite{AtrioBarandela:1996ur}, late-forming DM~\cite{Das:2006ht} and atomic DM~\cite{CyrRacine:2012fz}. Taking all these possibilities into account, it would be difficult to determine the specific nature of the DM coupling from this feature alone. Furthermore, since the oscillations are not as prominent as in the case of DM--photon interactions~\cite{Boehm:2001hm} or atomic DM in the sDAO (strong dark acoustic oscillation) scenario~\cite{Buckley:2014hja}, they may not be resolved. In this case, there could be a degeneracy with both warm DM~\cite{Viel:2005qj} and axion DM~\cite{Hlozek:2014lca} models, which predict a sharp cut-off in the matter power spectrum at small scales.

\section{Current constraints}
\label{sec:currentconstraints}

To assess how powerful the constraints from future LSS surveys can be, we first derive the limits set by current CMB and galaxy clustering surveys. These will then serve as a benchmark for our forecasts in Sec.~\ref{sec:forecasts}.

To perform this analysis, the modifications shown in Eq.~(\ref{Eq:nudmequations}) are implemented in the Boltzmann code {\sc class}~\cite{Lesgourgues:2011re} (see also Ref.~\cite{Wilkinson:2014ksa}) and the posterior likelihoods are obtained using the Markov Chain Monte Carlo (MCMC) code {\sc Monte Python}~\cite{Audren:2012wb}. The prior ranges for these parameters are listed in Tab.~\ref{tab:priors}. Since $u$ can vary by many orders of magnitude, we select a logarithmic prior distribution for this parameter, in contrast to the linear priors used in Refs.~\cite{Mangano:2006mp,Serra:2009uu,Wilkinson:2014ksa}.

\begin{table}[t]
\begin{center}
\begin{tabular}{cc}
\hline\hline
Parameter & Prior\\
\hline
$\Omega_{\rm b} h^2$ & $0.005 \to 0.1$\\
$\Omega_{\rm DM} h^2$ & $0.01 \to 0.99$\\
$100\hspace{0.4ex}h$ & $50 \to 100$\\
$10^9 A_{\rm s}$ & $1 \to 4$\\
$n_{\rm s}$ & $0.5 \to 1.5$\\
$\tau_{\rm reio}$ & $0.01 \to 0.1$\\
log$(u)$ {\rm (s-wave)} & $-6 \to 0$\\
log$(u)$ {\rm (p-wave)} & $-18 \to -11$\\
\hline\hline
\end{tabular}
\caption{Flat priors for the cosmological parameters considered here. $\Omega_{\rm{b}}h^2$ is the baryon density, $\Omega_{\rm{DM}}h^2$ is the DM density, $h$ is the reduced Hubble parameter, $A_{\rm s}$ is the primordial spectrum amplitude, $n_{\rm s}$ is the scalar spectral index, $\tau_{\rm{reio}}$ is the optical depth and $u$ is defined in Eq.~\eqref{eq:u_nudm}.} 
\label{tab:priors}
\end{center}
\end{table}

For simplicity, we assume massless neutrinos\footnote{This is in contrast to \Planck, whose analysis assumes two massless and one massive neutrino with $m_{\nu} = 0.06$ eV~\cite{Ade:2013zuv}. Such a small neutrino mass only affects the CMB through a slight shift in the angular diameter distance, which can be exactly compensated by a decrease in $100\hspace{0.4ex}h$ of $\sim 0.6$~\cite{Ade:2013zuv} \label{mnu}.} and fix the effective number of neutrino species, $N_{\rm eff}$, to the standard value of 3.046~\cite{Mangano:2005cc}. We have verified that allowing $N_{\rm eff}$ to vary has an impact on the value of the Hubble parameter, $H_0$, but does not change the sensitivity to the $u$ parameter.

The current CMB constraints (using \Planck 2013 + WMAP polarisation data~\cite{Ade:2013zuv}) are shown in Tab.~\ref{Table:nudmconstraints}. The corresponding upper limits on the DM--neutrino scattering cross section (at 95\% CL) are
\begin{equation}
\sigma_{{\rm DM}-\nu,0}^{(\Planck)} \lesssim 6 \times 10^{-31} \left(m_{\rm{DM}}/\rm{GeV}\right) \ \rm{cm^2}~,
\end{equation}
if s-wave and  
\begin{equation}
\sigma_{{\rm DM}-\nu,2}^{(\Planck)} \lesssim 2 \times 10^{-40}  \left(m_{\rm{DM}}/\rm{GeV}\right) \ \rm{cm^2}~,
\end{equation}
if p-wave. These results are consistent with those quoted by the authors of Refs.~\cite{Mangano:2006mp,Serra:2009uu}, with the caveat that they did not perform a full MCMC analysis.

\begin{table*} 
\begin{tabular*}{\textwidth}{{@{\extracolsep{\fill} }c|ccc|ccc}} 
\toprule
& \multicolumn{2}{c}{s-wave ($u = {\rm const.}$)} &&  \multicolumn{2}{c}{p-wave ($u \propto T^2$)}  \\ \hline
Parameter &\Planck 2013 & COrE+ && \Planck 2013 & COrE+ \\ \hline 
$\Omega_{\rm b} h^2$& $0.0221\pm 0.0003$ &  $0.02223\pm 0.00004$ && $0.0221\pm 0.0003$ &  $0.02222\pm 0.00004$\\ 
$\Omega_{\rm DM} h^2$  & $0.120 \pm 0.003$  & $0.1199 \pm 0.0005$&& $0.119 \pm 0.003$ &  $0.1197\pm 0.0005$ \\ 
$100\hspace{0.4ex}h$ & $68.0 \pm 1.3 $ & $67.3 \pm 0.2 $  && $68.0 \pm 1.2 $ & $67.3\pm 0.2 $\\ 
$10^{9}A_{\rm s}$ & $2.20 \pm 0.06 $ &  $2.207 \pm 0.010 $  &&  $2.19 \pm 0.06 $  &  $2.207 \pm 0.010 $\\ 
$n_{\rm s}$ & $0.961 \pm 0.008$ &  $0.9656 \pm 0.0017$ && $0.961 \pm 0.008$  &  $0.9639\pm 0.0019$\\ 
$\tau_{\rm reio }$  & $0.090\pm 0.015 $&  $0.0792\pm 0.0002 $ && $0.090\pm 0.013 $ &   $0.0790 \pm 0.0002$ \\ 
$\log_{10}(u)$ & $<-4.04$ (95\% CL) &  $-4.33$ (95\% CL) && $<-13.6$ (95\% CL) & $ <-14.6$ (95\% CL)  \\ 
\hline \hline 
\end{tabular*}
\caption{Marginalised posteriors for s-wave (left) and p-wave (right) DM--neutrino scattering cross sections set by the \Planck 2013 data (+ WMAP polarisation) (see Sec.~\ref{sec:currentconstraints}) and the COrE+ forecast (see Sec.~\ref{sec:forecasts}). Unless otherwise indicated, the errors represent the 68\% CL.}
\label{Table:nudmconstraints}
\end{table*} 

\begin{figure*}
\begin{tabular}{c c}
\includegraphics[width=0.48\textwidth]{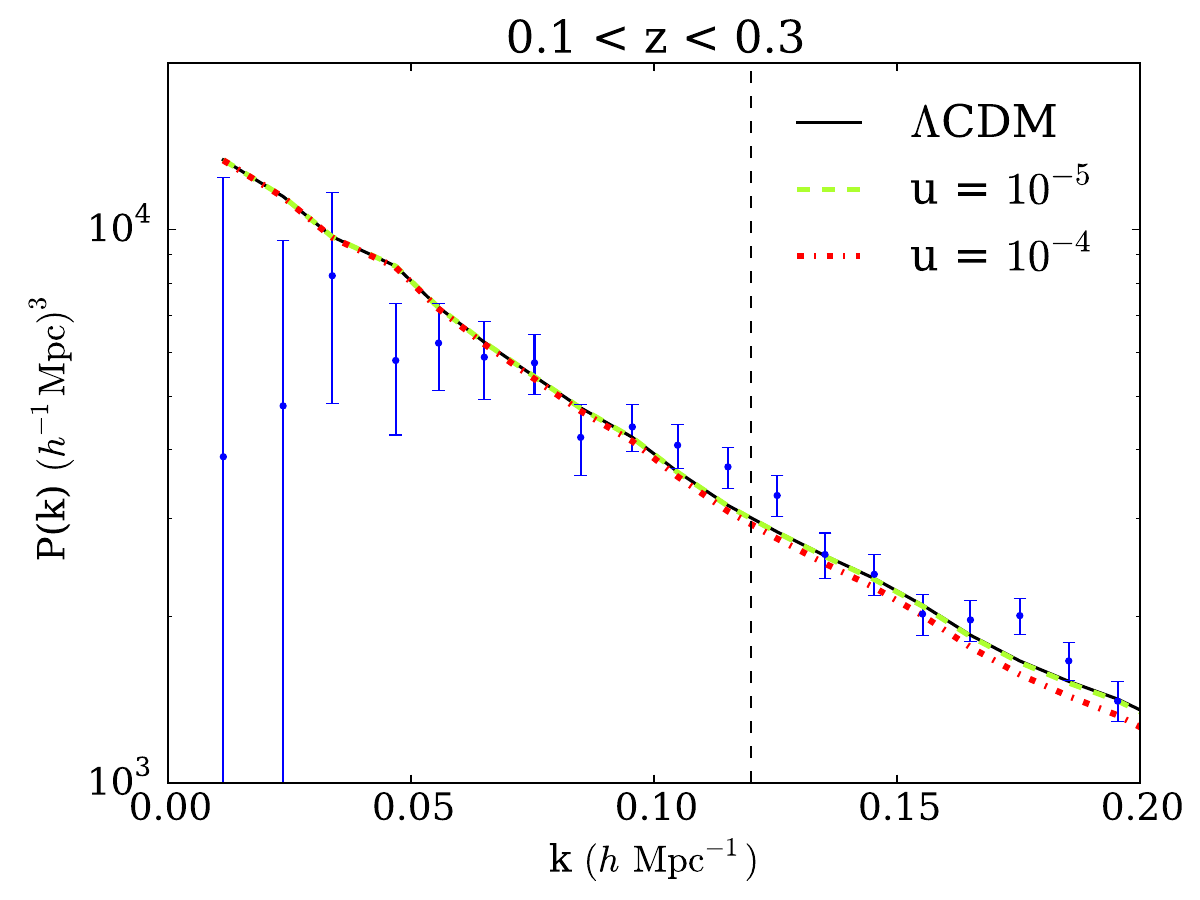} &\includegraphics[width=0.48\textwidth]{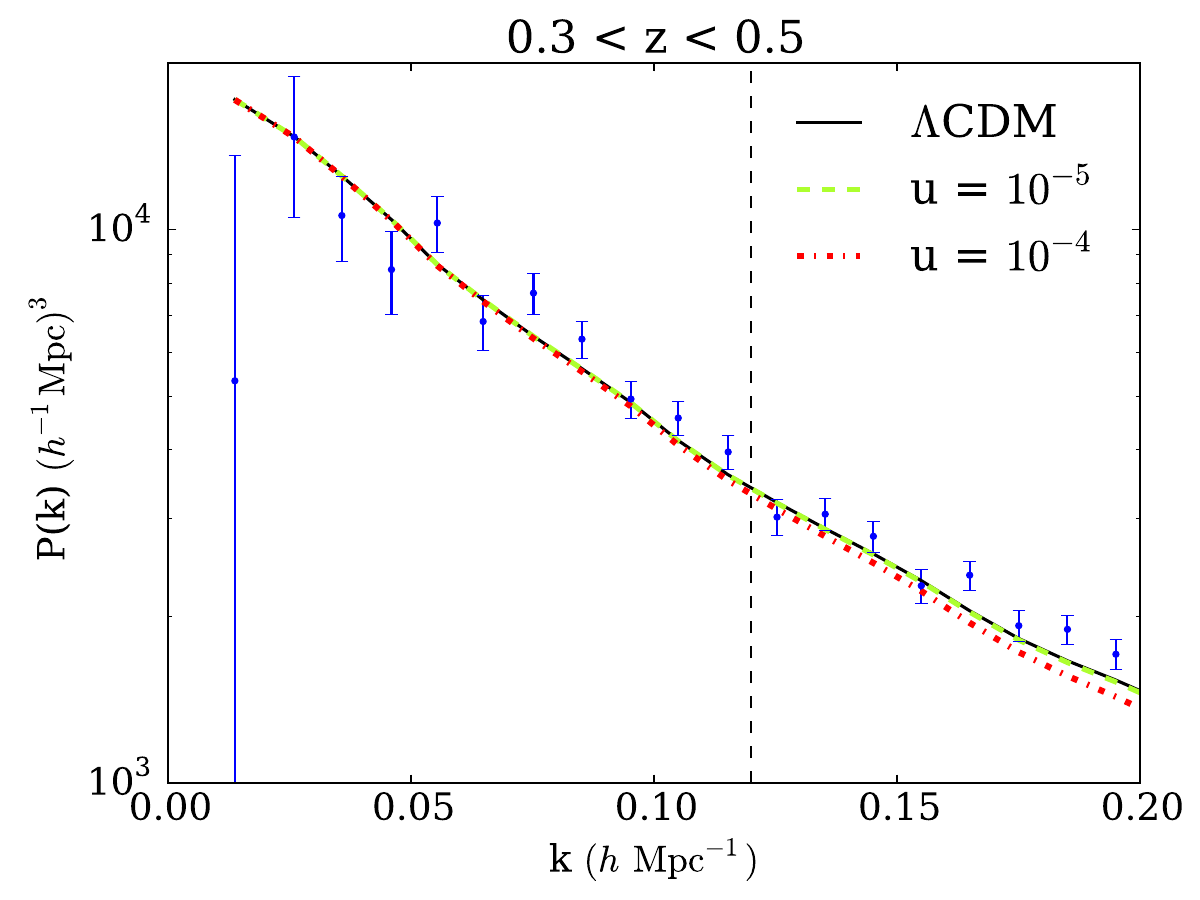}\\
\includegraphics[width=0.48\textwidth]{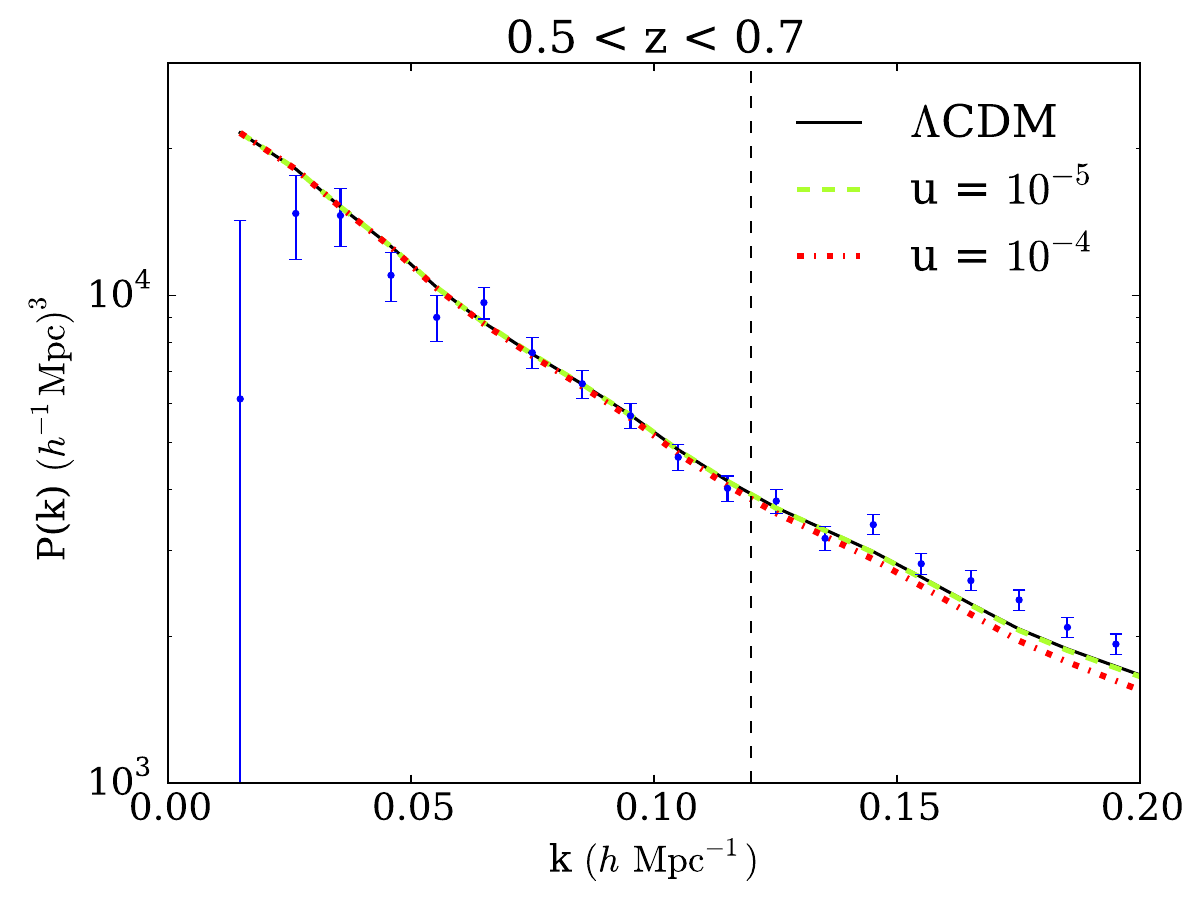} &\includegraphics[width=0.48\textwidth]{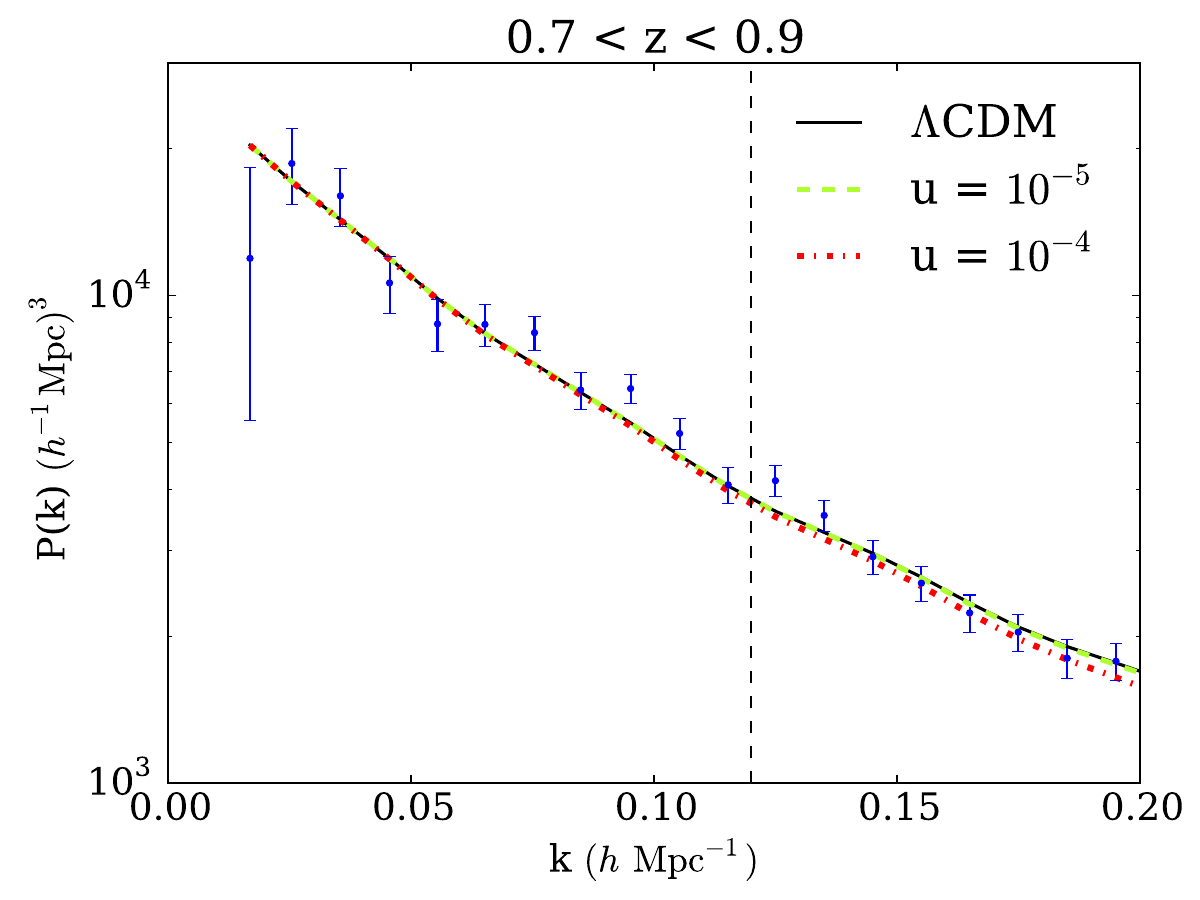}\\
\end{tabular}
 \caption{The data points show the galaxy power spectrum, $\hat{P}_{{\rm halo}}(k)$, in the four redshift bins ($0.1<z<0.3$, $0.3<z<0.5$, $0.5<z<0.7$ and $0.7<z<0.9$) measured by the \WiggleZ survey~\cite{Parkinson:2012vd}. We also depict the convolved power spectrum, $P_{{\rm halo},w}(k_i,\vartheta_\alpha)$, as defined in Eq.~(\ref{Eq:Phalowindef}), for the \lcdm fiducial model of Ref.~\cite{RiemerSorensen:2011fe} (solid black) and for two values of the $u$ parameter: $u = 10^{-4}$ (dotted red) and $u = 10^{-5}$ (dashed green) in the s-wave scenario. The vertical dashed line denotes the separation between the linear ($k \lesssim 0.12 \ h$~Mpc$^{-1}$) and non-linear ($k \gtrsim 0.12 \ h$~Mpc$^{-1}$) regimes.}
\label{Fig:WiggleZ}
\end{figure*}

We now repeat the previous analysis adding LSS data on the full shape of the matter power spectrum. Concretely, we use the galaxy clustering information from the \WiggleZ Dark Energy Survey~\cite{Parkinson:2012vd}. The \WiggleZ sample consists of $\sim 238,000$ galaxies and covers a region of 1 ${\rm Gpc}^3$ in redshift space. Our calculations have shown that comparable results can be obtained from the BOSS DR11 measurements~\cite{Anderson:2013zyy}. Following a similar analysis to Ref.~\cite{Reid:2009xm}, we construct the likelihood function as follows:
\begin{equation}
-2 \ {\rm log}[L(\vartheta_\alpha)] = \chi^2 (\vartheta_\alpha) = \sum_{ij} \Delta_i C^{-1} _{ij} \Delta_j~,
\label{Eq:MPSchi2}
\end{equation}
where the covariance matrix reads
\begin{equation}
C_{ij} = \langle \hat{P}_{{\rm halo}}(k_i) \hat{P}_{{\rm halo}}(k_j) \rangle - \langle \hat{P}_{{\rm halo}}(k_i)\rangle \langle \hat{P}_{{\rm halo}}(k_j)\rangle~,
\label{Eq:MPScov}
\end{equation}
and
\begin{equation}
 \Delta_i \equiv \left[ \hat{P}_{{\rm halo}}(k_i)-P_{{\rm halo},w}(k_i,\vartheta_\alpha) \right]~.
\label{Eq:DeltaMPS}
\end{equation}
In Eq.~\eqref{Eq:DeltaMPS}, $\hat{P}_{{\rm halo}}(k_i)$ is the measured galaxy power spectrum and $P_{{\rm halo},w}(k_i,\vartheta_\alpha)$ is the theoretical expectation for the set of model parameters $\vartheta_\alpha$, listed in Tab.~\ref{tab:priors}. In turn, $P_{{\rm halo},w}(k_i,\vartheta_\alpha)$ is a convolution of the computed galaxy power spectrum with the survey window functions, $W(k_i,k_n)$, and is given by
\begin{equation}
P_{{\rm halo},w}(k_i,\vartheta_\alpha) = \sum_n \frac{W(k_i,k_n)P_{{\rm halo}}(k_n/a_{\rm scl}, \vartheta_\alpha)}{a^3_{\rm scl}}~.
\label{Eq:Phalowindef}
\end{equation}
In this equation, $a_{\rm scl}$ represents the scaling, which takes into account that the observed galaxy redshift has to be translated into a distance using a fiducial model. In this case, we use the same values as in Ref.~\cite{RiemerSorensen:2011fe}: $\Omega_{\rm b} = 0.049$, $\Omega_{\rm m} = 0.297$, $h = 0.7$, $n_{\rm s} = 1$ and $\sigma_8 = 0.8$. The scaling factor is given by Refs.~\cite{Tegmark:2006az,Reid:2009xm}:
\begin{equation}
a^3_{\rm scl} = \frac{D_A(z)^2 H(z)}{D_{A,{\rm fid}}(z)^2 H_{\rm fid}(z)}~.
\label{Eq:ascaling}
\end{equation}
The theoretical galaxy power spectrum $P_{{\rm halo}} (k,\vartheta_\alpha)$ is related to the matter power spectrum $P^{m}(k,\vartheta_\alpha)$ through the relation 
\begin{equation}
P_{{\rm halo}} (k,\vartheta_\alpha)=b^2 \ P^{m} (k,\vartheta_\alpha)~,
\label{Eq:Ptheo}
\end{equation}
where $b$ is the bias, which is assumed to be constant. We analytically marginalise over $b$ as in Ref.~\cite{Lewis:2002ah}:
\begin{equation}
b^2 = \frac{\sum_{i j} P_{{\rm halo},w}(k_i, \vartheta_\alpha) C^{-1} _{i j} \hat{P}_{{\rm halo}}(k_j)}{\sum_{i j} P_{{\rm halo},w}(k_i, \vartheta_\alpha) C^{-1} _{i j} P_{{\rm halo},w}(k_j,\vartheta_\alpha)}~.
\label{Eq:biasmargina}
\end{equation}
In Fig.~\ref{Fig:WiggleZ}, we show the measured galaxy power spectrum, $\hat{P}_{{\rm halo}}(k)$, from \WiggleZ in the four redshift bins ($0.1<z<0.3$, $0.3<z<0.5$, $0.5<z<0.7$ and $0.7<z<0.9$) exploited in our analyses~\cite{Parkinson:2012vd}. We also depict the convolved power spectrum, $P_{{\rm halo},w}(k_i,\vartheta_\alpha)$, as defined in Eq.~(\ref{Eq:Phalowindef}), for the \lcdm fiducial model of Ref.~\cite{RiemerSorensen:2011fe} and for two values of the $u$ parameter ($u=10^{-4}$ and $u=10^{-5}$, both in the s-wave scenario). The characteristic damping of the ${\rm P}(k)$ due to the interacting DM--neutrino fluid is clearly visible and allows us to tighten the constraints with respect to the previous CMB-only analysis.

\begin{table*} 
\begin{tabular*}{\textwidth}{{@{\extracolsep{\fill} }c|ccc|ccc}} 
\toprule
& \multicolumn{2}{c}{s-wave ($u = {\rm const.}$)} &&  \multicolumn{2}{c}{p-wave ($u \propto T^2$)}  \\ \hline
Parameter & $k_{\rm max} = 0.12 \ h$~Mpc$^{-1}$ & $k_{\rm max} = 0.2 \ h$~Mpc$^{-1}$   && $k_{\rm max} = 0.12 \ h$~Mpc$^{-1}$  & $k_{\rm max} = 0.2 \ h$~Mpc$^{-1}$  \\ \hline 
$\Omega_{\rm b} h^2$& $0.0220\pm 0.0003$ &  $0.0219\pm 0.0003$ && $0.0219\pm 0.0003$ &   $0.0218\pm 0.0003$\\ 
$\Omega_{\rm DM} h^2$  & $0.122 \pm 0.002$  & $0.123 \pm 0.003$ && $0.122 \pm 0.002$ &  $0.123 \pm 0.002$  \\ 
$100\hspace{0.4ex}h$ & $67.0 \pm 1.1 $ & $66.6 \pm 1.0 $  && $66.9 \pm 1.1 $ & $66.7 \pm 1.0 $\\ 
$10^{9}A_{\rm s}$ & $2.19 \pm 0.05 $ &  $2.19 \pm 0.05 $  &&  $2.19 \pm 0.05 $  &  $2.19 \pm 0.05 $\\ 
$n_{\rm s}$ & $0.956 \pm 0.007$ &  $0.956 \pm 0.006$ && $0.956 \pm 0.007$  &  $0.955 \pm 0.007$\\ 
$\tau_{\rm reio }$  & $0.086\pm 0.013 $&  $0.086\pm 0.013 $ && $0.085\pm 0.013 $ &   $0.085\pm 0.013 $ \\ 
$\log_{10}(u)$ & $<-4.18$ (95\% CL) &  $-4.57$ (95\% CL) && $<-13.7$ (95\% CL) & $ <-13.9$ (95\% CL)  \\ 
\hline \hline 
\end{tabular*}
\caption{Marginalised posteriors for s-wave (left) and p-wave (right) DM--neutrino scattering cross sections set by the combination of \WiggleZ full-shape galaxy power spectrum measurements and \Planck 2013 (+ WMAP polarisation) data. Unless otherwise indicated, the errors represent the 68\% CL.}
\label{Table:LSSconstraints}
\end{table*}

In Tab.~\ref{Table:LSSconstraints}, we present the posteriors obtained using the combination of \WiggleZ and CMB data. We perform two separate analyses, including data for which: (i)  $k<k_{\textrm{max}}=0.12 \ h$~Mpc$^{-1}$ (purely linear regime) and (ii) $k<k_{\textrm{max}}=0.2 \ h$~Mpc$^{-1}$ (weakly non-linear regime).

In terms of the DM--neutrino scattering cross section (at 95\% CL) with $k_{\textrm{max}}=0.12 \ h$~Mpc$^{-1}$  ($k_{\textrm{max}}=0.2 \ h$~Mpc$^{-1}$), we obtain
\begin{eqnarray}
\sigma_{{\rm DM}-\nu,0}^{(\WiggleZ)} &\lesssim& 4 \times 10^{-31} \left(m_{\rm{DM}}/\rm{GeV}\right) \ \rm{cm^2}~;\nonumber\\
(&\lesssim &2 \times 10^{-31} \left(m_{\rm{DM}}/\rm{GeV}\right) \ \rm{cm^2})~,
\label{eq:sigmalss}
\end{eqnarray}
for the s-wave cross section. As we shall see in the next section, these bounds are competitive with those resulting from our forecasts for the future CMB mission COrE+. 

Meanwhile, for the p-wave cross section, we obtain
\begin{eqnarray}
\sigma_{{\rm DM}-\nu,2}^{(\WiggleZ)} &\lesssim& 1 \times 10^{-40} \left(m_{\rm{DM}}/\rm{GeV}\right) \ \rm{cm^2}~;\nonumber\\
(&\lesssim &8 \times 10^{-41} \left(m_{\rm{DM}}/\rm{GeV}\right) \ \rm{cm^2})~.
\end{eqnarray}
Therefore, including data in the weakly non-linear regime ($k < 0.2 \ h$~Mpc$^{-1}$) only strengthens the constraints by a factor of $ 2$ (s-wave) and $1.25$ (p-wave) with respect to those in the purely linear regime ($k < 0.12 \ h$~Mpc$^{-1}$). We note that, in the s-wave scenario, the bounds are as much as $\sim 3.5$ times tighter than those using only CMB measurements, showing the benefits of utilising the full shape of the ${\rm P}(k)$. The improvement is not as significant for the p-wave case because the suppression appears at larger scales (see e.g. Ref.~\cite{Blennow:2012de}).

\section{Forecasts for future experiments}
\label{sec:forecasts}

The CMB and LSS analyses in Sec.~\ref{sec:currentconstraints} allowed us to obtain current constraints on the DM--neutrino elastic scattering cross section. We will now assess the power of future experiments in (i) \textit{constraining} DM microphysics and (ii) \textit{detecting} small deviations from the \lcdm matter power spectrum in the weakly non-linear regime. These two analyses require slightly different methodologies. In the first case, we construct a mock catalogue based on the \lcdm cosmology and compute the strongest possible upper limit on the $u$ parameter using the expected sensitivity of future experiments. In the second case, the mock data assumes small but non-negligible DM--neutrino interactions in order to assess our ability to detect them and more generally, reconstruct possible deviations from \lcdm. In both cases, we use projected sensitivities.

As in the previous section, we first consider CMB observables only and then include data from LSS surveys. We focus on two planned experiments: (i) COrE+~\cite{Bouchet:2011ck}, a CMB space mission currently proposed for the 2015-2025 ESA call, and (ii) DESI~\cite{Levi:2013gra}, a multiplexed fibre-fed spectrograph to detect galaxies and quasars up to redshift $z \sim 2$, that is expected to run in the 2018-2022 timeframe.

\subsection{COrE+}
\label{subsec:COrE+}

We first produce full mock CMB data sets (temperature and $E$-polarisation, plus lensing). We then compute the fiducial angular power spectra, $C_\ell$, using the best-fit cosmology reported by the \Planck 2015 final mission, including the TT, TE and EE spectra~\cite{Adam:2015rua}. To these $C_\ell$, we add a noise component $N_\ell$ consistent with each COrE+ channel specification and given by 
\begin{equation}\label{Eq:CMBforecastnoise}
N_\ell^{IJ} =\delta_{IJ} \, \sigma^I\sigma^J \exp \left[ \ell \left( \ell+1 \right) \frac{\theta^2}{8\hspace{0.4ex}{\rm ln}2} \right]~,
\end{equation}
where $\sigma^{I,J}$ correspond to the temperature or polarisation errors (i.e. $I,J \in$ \{T,E\}). The expected temperature and polarisation sensitivities are given in Tab.~\ref{Table:Noise}.

Following Ref.~\cite{Perotto:2006rj}, the effective $\chi^2$ is given by
\begin{equation}
\label{Eq:CMBforecastchi}
\chi^2_{\rm eff} (\vartheta_\alpha) = \sum_\ell (2 \ell +1) f_{\rm sky} \left( \frac{D}{|\bar{C}|} + {\rm ln} \frac{|\bar{C}|}{|\hat{C}|} - 3 \right)~,
\end{equation}
where $D$ is a certain function of the noised power spectra (see Eq.~(3.4) in Ref.~\cite{Perotto:2006rj}) and $|\bar{C}|$ and $|\hat{C}|$ represent the determinants of the theoretical and observed covariance matrices respectively. Finally, $f_{\rm sky}$ represents the observed fraction of the sky (in practice, it weights the correlations between multipoles when the map does not cover the full sky). For this analysis, we use $f_{\rm sky} = 0.7$~\cite{Bouchet:2011ck}.

 \begin{table}
\begin{tabular}{cccc} 
  \toprule
 Channel & $\theta$ &  $\Delta$T & $\Delta$P  \\
(GHz) & (arcmin) &   ($\mu {\rm K} \cdot$arcmin) & ($\mu {\rm K} \cdot$arcmin) \\
\hline
 105 & 10.0 & 2.68 & 4.63 \\
 135 & 7.8 & 2.63 & 4.55 \\ 
 165 & 6.4 & 2.67 & 4.61 \\ 
 195 & 5.4 & 2.63 & 4.54 \\   \toprule
 \end{tabular}
\caption{COrE+ 4-year sensitivity. $\theta$ is the Full Width at Half Maximum (FWHM) of the beam, $\Delta$T and $\Delta$P are the temperature and polarisation sensitivities respectively~\cite{Bouchet:2011ck}.}
 \label{Table:Noise} 
\end{table}

The third step in our analysis is to compute a Gaussian likelihood around our fiducial spectra, using {\sc class}, tuned to obtain a 0.01\% precision on the $C_\ell$ (as in Ref.~\cite{Perotto:2006rj}, according to Eq.~\eqref{Eq:CMBforecastchi} and with a noise given by Eq.~\eqref{Eq:CMBforecastnoise}). Then, assuming a 4-year sensitivity and using {\sc Monte Python} to sample the parameter space with the priors given in Tab.~\ref{tab:priors}, we can predict the sensitivity of COrE+ to the \lcdm parameters. Note that we only consider the TT, TE and EE observables. For simplicity, we neglect tensor modes (i.e. BT, BE and BB) as they have currently not been observed~\cite{Ade:2015lrj}.

The results are presented in Tab.~\ref{Table:nudmconstraints}. We infer that the future sensitivity of COrE+ to a DM--neutrino coupling would be (at 95\% CL)
\begin{equation}
\sigma_{{\rm DM}-\nu,0}^{(\rm COrE+)} \lesssim 3 \times 10^{-31} \left(m_{\rm{DM}}/\rm{GeV}\right) \ \rm{cm^2}~,
\label{Eq:coreforecastswave}
\end{equation}
if s-wave and
\begin{equation}
\sigma_{{\rm DM}-\nu,2}^{(\rm COrE+)} \lesssim 2 \times 10^{-41} \left(m_{\rm{DM}}/\rm{GeV}\right) \ \rm{cm^2}~,
\end{equation}
if p-wave.

While we find that the standard cosmological parameters will be measured to much higher precision than with \Planck, there is only a modest gain in sensitivity to the DM--neutrino cross section. Furthermore, these limits are slightly worse than those obtained after combining \Planck with current LSS data in the weakly non-linear regime. From these results, we expect detection with COrE+ to be possible for $u \simeq 10^{-4}$.

To assess the power of COrE+ to detect and reconstruct the \nudm cosmology or similar deviations to \lcdm, we also produce mock data sets with $u = 10^{-4}$ and $u = 10^{-5}$ as fiducial models (for s-wave interactions). We then attempt to reconstruct these models by means of the usual MCMC method. The $u = 10^{-5}$ case is presented in Fig.~\ref{fig:uom} (and similarly for p-wave with $u = 10^{-14}$). With COrE+--like CMB data, one may reconstruct a universe with $u = 10^{-4}$ with a $40\%$ $1 \sigma$ error. However, the $u = 10^{-5}$ case would provide us with CMB information entirely consistent with $u = 0$, in agreement with Eq.~\eqref{Eq:coreforecastswave}. Therefore, $u \gtrsim 5\times10^{-5}$ is the best sensitivity that one could achieve with CMB experiments in the near future.

\subsection{DESI}
\label{subsec:DESI}

The DESI survey~\cite{Levi:2013gra} is expected to provide a wealth of information on the matter distribution (i.e. the P$(k)$) in the Universe at relatively small scales and up to redshift $z \sim 2$. To forecast the ability of DESI to discover new physics, we first compute the expected errors from the DESI instrument, following a Fisher matrix approach, which is the usual method used to forecast galaxy survey experiments\footnote{\url{http://desi.lbl.gov/}}${}^{,}$\footnote{\url{http://sci.esa.int/euclid/}}.

The Fisher matrix is defined as the expectation value of the second derivative of the likelihood surface around its maximum. As long as the posterior distribution for the parameters is well approximated by a multivariate Gaussian function, its elements are given by~\cite{Tegmark:1996bz,Jungman:1995bz,Fisher:1935e}
\begin{equation}
\label{eq:fish}
F_{\alpha\beta}=\frac{1}{2}{\rm
  Tr}\left[C^{-1}\frac{\partial C}{\partial \vartheta_\alpha}C^{-1}\frac{\partial C}{\partial \vartheta_\beta}\right]~,
\end{equation}
where $C=S+N$ is the total covariance, which consists of signal $S$ and noise $N$ terms. Once more, we take a fiducial cosmology defined by the parameters that best fit the \textit{Planck} 2015 TT, TE, EE $+$ lowP data~\cite{Planck:2015xua} in the presence of DM--neutrino interactions with $u = 10^{-5}$ in the s-wave scenario and $u = 10^{-14}$ in the p-wave scenario\footnote{See Ref.~\cite{Seo:2003pu} for more details on the Fisher matrix formalism for galaxy redshift surveys such as DESI.}.

Assuming a Gaussian likelihood for the DESI band powers, the Fisher matrix can be written as:
\begin{eqnarray}  
F^{\rm LSS}_{\alpha \beta}&=&\int_{\vec{k}_{\rm min}} ^ {\vec{k}_{\rm max}} \frac{\partial \ln P_{\rm gg}(\vec{k})}{\partial \vartheta_\alpha} \frac{\partial \ln P_{\rm gg}(\vec{k})}{\partial \vartheta_\beta} V_{\rm eff}(\vec{k}) \frac{d\vec{k}}{2(2 \pi)^3}  
\label{eq:Fij} \\
&=&\int_{-1}^{1} \int_{k_{\rm min}}^{k_{\rm max}}\frac{\partial \ln P_{\rm gg}(k,\mu)}{\partial \vartheta_\alpha} \frac{\partial \ln P_{\rm gg}(k,\mu)}{\partial \vartheta_\beta}  V_{\rm{eff}}(k,\mu)\nonumber\\
& &\frac{2\pi k^2 dk d\mu}{2(2 \pi)^3}~,  \nonumber 
\end{eqnarray}
where $V_{\rm eff}$ is the effective volume of the survey and given by 
\begin{eqnarray}
V_{\rm{eff}}(k,\mu) &=&\left [ \frac{{n}P(k,\mu)}{{n}P(k,\mu)+1} \right ]^2 V_{\textrm{survey}}~,
\label{eq:Veff} 
\end{eqnarray}
 where $\mu$ is the cosine of the angle between the vector mode ($\vec{k}$) and the vector along the line of sight, and
$n$ is the galaxy number density (which is assumed to be constant throughout each of the redshift bins).

To perform the analysis, we divide the data in redshift bins of width $\Delta z = 0.1$ and cut the small-scale data at $k =  0.25 \ h$~Mpc$^{-1}$ to avoid the highly non-linear regime. The lowest wavenumber (i.e. the largest scale), $k_{\rm min}$, is chosen to be greater than $2\pi/\Delta V^{1/3}$, where $\Delta V$ represents the volume of the redshift shell. We note that using data in the non-linear regime would require numerical simulations of this model. This has been performed for specific cases in Refs.~\cite{Boehm:2014vja,Schewtschenko:2014fca}. As we will discuss later, constraints using this method are competitive with our DESI forecast.

The real-space linear DM power spectrum, $P_{\rm DM}$, is related to the linear redshift-space galaxy power spectrum, $P_{\rm gg}$, by
\begin{equation}
P_{\rm gg}(k)=P_{\rm DM}(k) \ (b \ + \ \beta  \ \mu^2)^2~,
\label{eq:pspectrum}
\end{equation} 
where $b$ is the bias relating galaxy to DM overdensities in real space (as in Eq.~(\ref{Eq:Ptheo})) and $\beta$ is the linear growth factor.

DESI is expected to cover 14,000 deg$^2$ of the sky in the redshift range $0.15<z<1.85$. We use the values of the bias given in Ref.~\cite{Font-Ribera:2013rwa} for the three types of DESI tracers, namely $b_{\rm ELG}(z) D(z) = 0.84$ for the Emission Line Galaxies (ELGs), $b_{\rm LRG}(z) D(z) = 1.7$ for the Luminous Red Galaxies (LRGs) and $b_{\rm QSO}(z) D(z) = 1.2 $ for the high redshift quasars (QSOs). Here, $D(z)$ is the normalised growth factor and both the bias and the growth factor are assumed to vary in each redshift bin accordingly to these expressions. To combine the Fisher matrices from the three DESI tracers, we use the multi-tracer technique of Ref.~\cite{Abramo:2013awa}.
 
For the s-wave scenario, we obtain a $1\sigma$ error on the $u$ parameter of
\begin{equation}
\delta u^{\rm (DESI)} \simeq 3.7 \times 10^{-6}~,
\end{equation}
for the fiducial value of $u = 10^{-5}$. For p-wave, we obtain:
\begin{equation}
\delta u^{\rm (DESI)} \simeq 4.4 \times 10^{-15}~,
\end{equation}
for the fiducial value of $u = 10^{-14}$. Crucially, DESI will ensure a $\sim 2.5 \sigma$ detection of DM--neutrino interactions if the strength of such a coupling is $u \simeq 10^{-5}$ (or a $\sim 2 \sigma$ detection for $u \simeq 10^{-14}$ if p-wave).

\begin{figure*}
\begin{centering}
\includegraphics[width=0.48\textwidth]{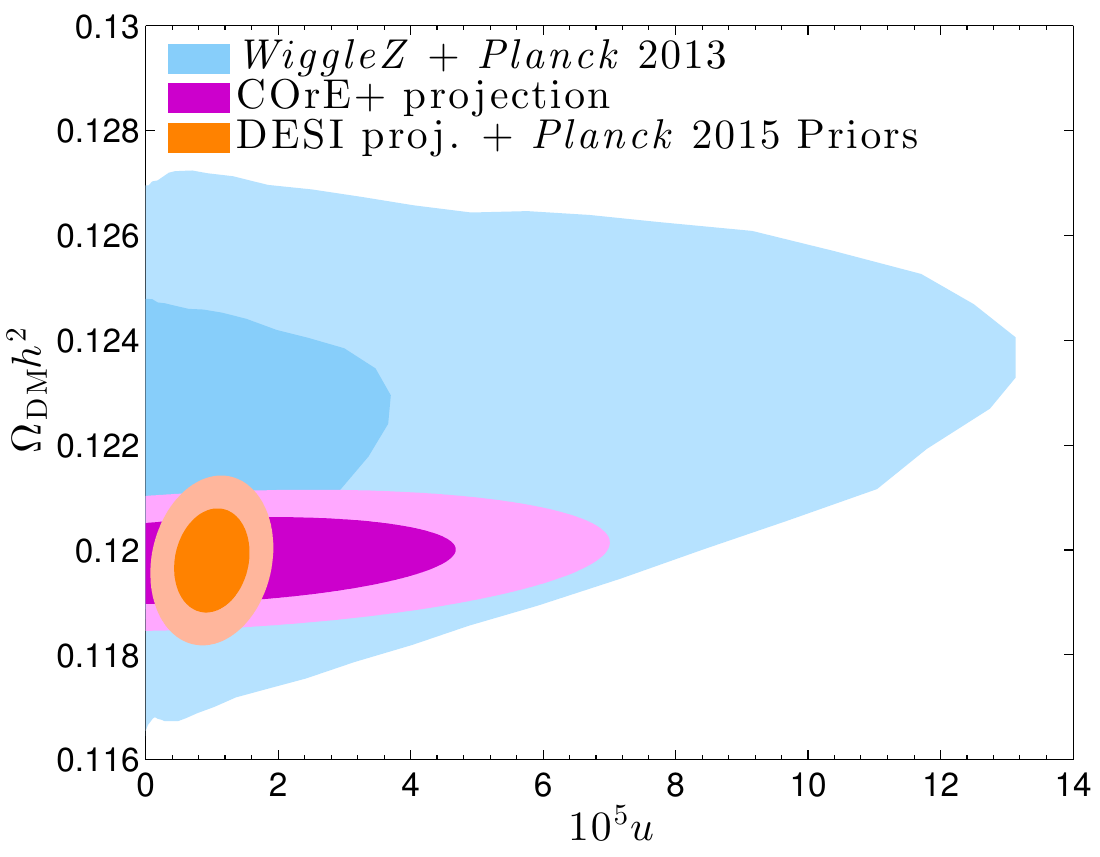} \hspace{1ex}
\includegraphics[width=0.48\textwidth]{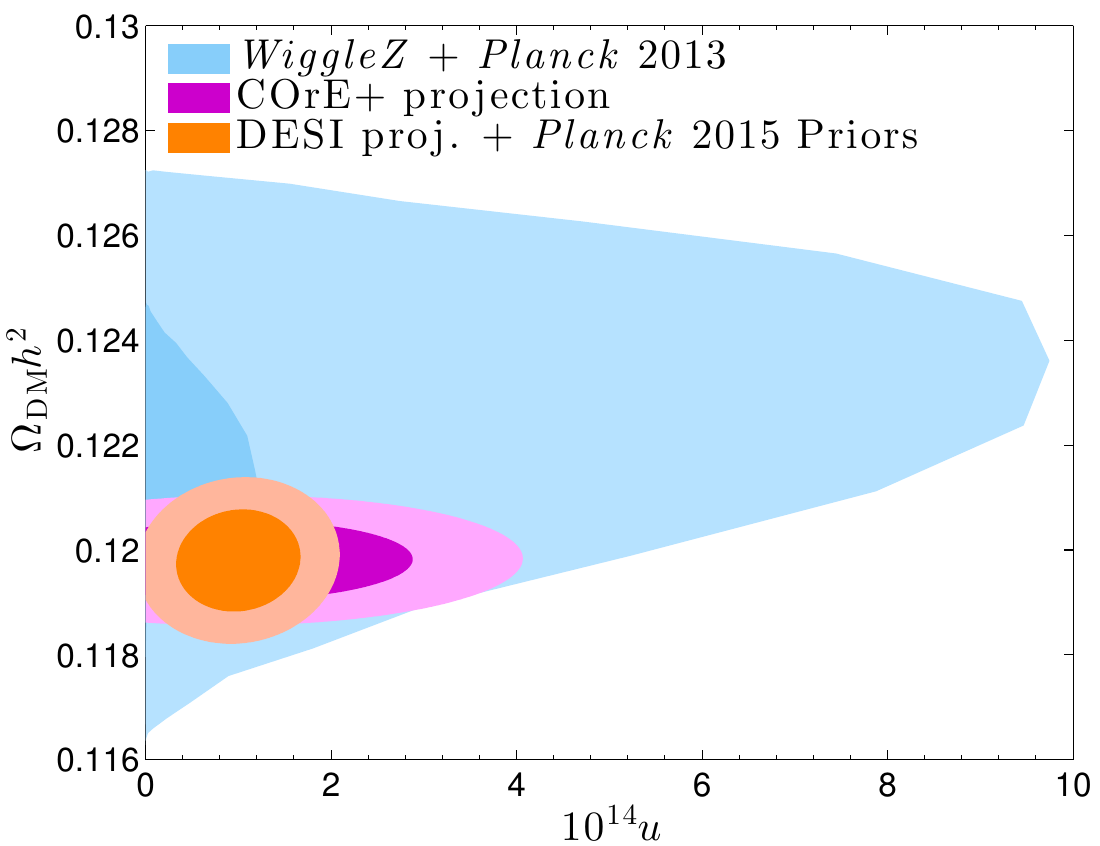}
\caption{The $68\%$ and $95\%$~CL allowed regions in the ($\Omega_{\rm{DM}}h^2$,$u$) plane for the s-wave (left) and p-wave (right) scenarios. Blue: current constraints from the combination of \WiggleZ and \Planck 2013 data, with $k_{\textrm{max}}=0.12 \ h$~Mpc$^{-1}$; Magenta: projected sensitivity of the upcoming COrE+ CMB experiment, assuming $u = 10^{-5}$ (or $u= 10^{-14}$ if p-wave); Orange: projected sensitivity of the DESI galaxy survey, again assuming $u = 10^{-5}$ (or $u = 10^{-14}$ if p-wave), with $k_{\textrm{max}}=0.25 \ h$~Mpc$^{-1}$.}
\label{fig:uom}
\end{centering}
\end{figure*}

Our results are summarised in Fig.~\ref{fig:uom}, where the DESI $68\%$ and $95\%$~CL allowed regions in the ($\Omega_{\rm{DM}}h^2$,$u$) plane are shown (assuming the \Planck 2015 fiducial cosmology plus an interaction strength of $u = 10^{-5}$ if s-wave and $u = 10^{-14}$ if p-wave), along with the current constraints and the COrE+ reconstruction. One can clearly see the improvement in the extraction of a DM--neutrino coupling that will be provided by the next-generation LSS surveys. This analysis indicates that planned galaxy clustering surveys will provide an extremely powerful tool (competitive or even better than future CMB experiments) to test the fundamental properties of DM.

Since the main impact of \nudm is the damping of structure on small scales, one of the largest effects will be a reduction in the number of satellites around galaxies such as the Milky Way. Until now, the only way to study interactions at these scales has been via $N$-body simulations, which show that for DM--radiation couplings greater than $u \simeq 10^{-5}$, the number of satellites in the Milky Way would be much smaller than observed~\cite{Boehm:2014vja,Schewtschenko:2014fca}. Therefore, with the sensitivity of $u \simeq 3.7 \times 10^{-6}$ expected from DESI, we would have a handle on alternative scenarios to \lcdm that modify our cosmic neighbourhood, independently of the assumptions that go into $N$-body simulations\footnote{With improvements in numerical algorithms and computing power, $N$-body simulations are becoming increasingly more affordable and will continue to provide a complementary method to test structure formation in models beyond \lcdm. However, they will remain computationally expensive, especially if one wishes to simulate structures on both large and small scales or test a wide range of modifications to the P$(k)$.}.

\section{Conclusion}
\label{sec:conc}

Cosmology provides a promising tool to measure the particle properties of dark matter (DM). A DM coupling to visible or dark radiation (including neutrinos, axions, dark photons or any other light uncharged particle) can lead to strong departures from the standard \lcdm cosmology and produce visible signatures for CMB experiments and LSS surveys. In the specific case of DM--neutrino scattering, one expects an enhancement of the CMB acoustic peaks due to the fact that DM is strongly coupled to neutrinos and vice versa, which delays the neutrino free-streaming epoch and alters DM clustering with respect to the standard \lcdm picture. However, the largest impact is imprinted as a damping in the matter power spectrum, surveyed by large-scale structure (LSS) galaxy surveys.

In this study, we have looked for the optimal method to measure such small departures from the \lcdm scenario. As cosmological measurements may constitute the only tool available to detect such effects, it is crucial to study the potential sensitivity of future experiments. We have shown that i) with current CMB measurements, one can probe s-wave and p-wave DM--neutrino cross sections of $\sigma_{{\rm DM}-\nu,0} \lesssim 6 \times 10^{-31} \left(m_{\rm{DM}}/\rm{GeV}\right) \ \rm{cm^2}$ and $\sigma_{{\rm DM}-\nu,2} \lesssim 2 \times 10^{-40} \left(m_{\rm{DM}}/\rm{GeV}\right) \ \rm{cm^2}$, respectively (at $95\%$~CL) and ii) by simulating a next-generation CMB experiment (i.e. a COrE+-like mission) by means of a Markov Chain Monte Carlo analysis, one can only weakly improve on the current sensitivity. 

The prospects for both constraints and detection are far better for future galaxy surveys, such as the DESI or Euclid experiments. Already, current LSS data, combined with \Planck CMB measurements, provide competitive constraints to those forecasted for a future CMB experiment such as COrE+. Future data from the DESI experiment alone could improve the current sensitivity limit by an order of magnitude, and provide an accurate (percent-level) measurement of the scattering cross section for values above that limit. Therefore, we have shown that galaxy clustering surveys are an excellent probe to detect new physics beyond \lcdm. Remarkably, future LSS experiments will be sensitive to effects that until now have only been accessible via $N$-body simulations.

\section{Acknowledgements}

The authors would like to thank Andreu Ari\~{n}o for useful discussions and Will Percival for precious BOSS DR11 measurements. The authors acknowledge the use of the publicly available numerical codes {\sc class} and {\sc Monte Python}, and the data from the \Planck Legacy Archive and \WiggleZ survey. ME is supported by Spanish grant FPU13/03111 of the MECD. RJW is supported by the STFC grant ST/K501979/1. This work was supported by the European Union FP7 ITN INVISIBLES (Marie Curie Actions, PITN-GA-2011-289442). We would like to thank the Universitad Aut\'onoma de Madrid and IFT for hospitality while this work was completed.

\bibliography{nudm_forecast.bib}

\begin{thebibliography}{64}%
\makeatletter
\providecommand \@ifxundefined [1]{%
 \@ifx{#1\undefined}
}%
\providecommand \@ifnum [1]{%
 \ifnum #1\expandafter \@firstoftwo
 \else \expandafter \@secondoftwo
 \fi
}%
\providecommand \@ifx [1]{%
 \ifx #1\expandafter \@firstoftwo
 \else \expandafter \@secondoftwo
 \fi
}%
\providecommand \natexlab [1]{#1}%
\providecommand \enquote  [1]{``#1''}%
\providecommand \bibnamefont  [1]{#1}%
\providecommand \bibfnamefont [1]{#1}%
\providecommand \citenamefont [1]{#1}%
\providecommand \href@noop [0]{\@secondoftwo}%
\providecommand \href [0]{\begingroup \@sanitize@url \@href}%
\providecommand \@href[1]{\@@startlink{#1}\@@href}%
\providecommand \@@href[1]{\endgroup#1\@@endlink}%
\providecommand \@sanitize@url [0]{\catcode `\\12\catcode `\$12\catcode
  `\&12\catcode `\#12\catcode `\^12\catcode `\_12\catcode `\%12\relax}%
\providecommand \@@startlink[1]{}%
\providecommand \@@endlink[0]{}%
\providecommand \url  [0]{\begingroup\@sanitize@url \@url }%
\providecommand \@url [1]{\endgroup\@href {#1}{\urlprefix }}%
\providecommand \urlprefix  [0]{URL }%
\providecommand \Eprint [0]{\href }%
\providecommand \doibase [0]{http://dx.doi.org/}%
\providecommand \selectlanguage [0]{\@gobble}%
\providecommand \bibinfo  [0]{\@secondoftwo}%
\providecommand \bibfield  [0]{\@secondoftwo}%
\providecommand \translation [1]{[#1]}%
\providecommand \BibitemOpen [0]{}%
\providecommand \bibitemStop [0]{}%
\providecommand \bibitemNoStop [0]{.\EOS\space}%
\providecommand \EOS [0]{\spacefactor3000\relax}%
\providecommand \BibitemShut  [1]{\csname bibitem#1\endcsname}%
\let\auto@bib@innerbib\@empty
\bibitem [{\citenamefont {Ade}\ \emph {et~al.}(2014)\citenamefont {Ade} \emph
  {et~al.}}]{Ade:2013zuv}%
  \BibitemOpen
  \bibfield  {author} {\bibinfo {author} {\bibfnamefont {P.}~\bibnamefont
  {Ade}} \emph {et~al.} (\bibinfo {collaboration} {Planck Collaboration}),\
  }\href {\doibase 10.1051/0004-6361/201321591} {\bibfield  {journal} {\bibinfo
   {journal} {Astron.Astrophys.}\ }\textbf {\bibinfo {volume} {571}},\ \bibinfo
  {pages} {A16} (\bibinfo {year} {2014})},\ \Eprint
  {http://arxiv.org/abs/1303.5076} {arXiv:1303.5076 [astro-ph.CO]} \BibitemShut
  {NoStop}%
\bibitem [{\citenamefont {Adam}\ \emph {et~al.}(2015)\citenamefont {Adam} \emph
  {et~al.}}]{Adam:2015rua}%
  \BibitemOpen
  \bibfield  {author} {\bibinfo {author} {\bibfnamefont {R.}~\bibnamefont
  {Adam}} \emph {et~al.} (\bibinfo {collaboration} {Planck Collaboration}),\
  }\href@noop {} {\  (\bibinfo {year} {2015})},\ \Eprint
  {http://arxiv.org/abs/1502.01582} {arXiv:1502.01582 [astro-ph.CO]}
  \BibitemShut {NoStop}%
\bibitem [{\citenamefont {Dodelson}\ and\ \citenamefont
  {Widrow}(1994)}]{Dodelson:1993je}%
  \BibitemOpen
  \bibfield  {author} {\bibinfo {author} {\bibfnamefont {S.}~\bibnamefont
  {Dodelson}}\ and\ \bibinfo {author} {\bibfnamefont {L.~M.}\ \bibnamefont
  {Widrow}},\ }\href {\doibase 10.1103/PhysRevLett.72.17} {\bibfield  {journal}
  {\bibinfo  {journal} {Phys. Rev. Lett.}\ }\textbf {\bibinfo {volume} {72}},\
  \bibinfo {pages} {17} (\bibinfo {year} {1994})},\ \Eprint
  {http://arxiv.org/abs/hep-ph/9303287} {arXiv:hep-ph/9303287 [hep-ph]}
  \BibitemShut {NoStop}%
\bibitem [{\citenamefont {Lesgourgues}\ and\ \citenamefont
  {Pastor}(2006)}]{Lesgourgues:2006nd}%
  \BibitemOpen
  \bibfield  {author} {\bibinfo {author} {\bibfnamefont {J.}~\bibnamefont
  {Lesgourgues}}\ and\ \bibinfo {author} {\bibfnamefont {S.}~\bibnamefont
  {Pastor}},\ }\href {\doibase 10.1016/j.physrep.2006.04.001} {\bibfield
  {journal} {\bibinfo  {journal} {Phys.Rept.}\ }\textbf {\bibinfo {volume}
  {429}},\ \bibinfo {pages} {307} (\bibinfo {year} {2006})},\ \Eprint
  {http://arxiv.org/abs/astro-ph/0603494} {arXiv:astro-ph/0603494 [astro-ph]}
  \BibitemShut {NoStop}%
\bibitem [{\citenamefont {Boehm}\ \emph {et~al.}(2001)\citenamefont {Boehm},
  \citenamefont {Fayet},\ and\ \citenamefont {Schaeffer}}]{Boehm:2000gq}%
  \BibitemOpen
  \bibfield  {author} {\bibinfo {author} {\bibfnamefont {C.}~\bibnamefont
  {Boehm}}, \bibinfo {author} {\bibfnamefont {P.}~\bibnamefont {Fayet}}, \ and\
  \bibinfo {author} {\bibfnamefont {R.}~\bibnamefont {Schaeffer}},\ }\href
  {\doibase 10.1016/S0370-2693(01)01060-7} {\bibfield  {journal} {\bibinfo
  {journal} {Phys.Lett.}\ }\textbf {\bibinfo {volume} {B518}},\ \bibinfo
  {pages} {8} (\bibinfo {year} {2001})},\ \Eprint
  {http://arxiv.org/abs/astro-ph/0012504} {arXiv:astro-ph/0012504 [astro-ph]}
  \BibitemShut {NoStop}%
\bibitem [{\citenamefont {Boehm}\ \emph {et~al.}(2002)\citenamefont {Boehm},
  \citenamefont {Riazuelo}, \citenamefont {Hansen},\ and\ \citenamefont
  {Schaeffer}}]{Boehm:2001hm}%
  \BibitemOpen
  \bibfield  {author} {\bibinfo {author} {\bibfnamefont {C.}~\bibnamefont
  {Boehm}}, \bibinfo {author} {\bibfnamefont {A.}~\bibnamefont {Riazuelo}},
  \bibinfo {author} {\bibfnamefont {S.~H.}\ \bibnamefont {Hansen}}, \ and\
  \bibinfo {author} {\bibfnamefont {R.}~\bibnamefont {Schaeffer}},\ }\href
  {\doibase 10.1103/PhysRevD.66.083505} {\bibfield  {journal} {\bibinfo
  {journal} {Phys.Rev.}\ }\textbf {\bibinfo {volume} {D66}},\ \bibinfo {pages}
  {083505} (\bibinfo {year} {2002})},\ \Eprint
  {http://arxiv.org/abs/astro-ph/0112522} {arXiv:astro-ph/0112522 [astro-ph]}
  \BibitemShut {NoStop}%
\bibitem [{\citenamefont {Boehm}\ and\ \citenamefont
  {Schaeffer}(2005)}]{Boehm:2004th}%
  \BibitemOpen
  \bibfield  {author} {\bibinfo {author} {\bibfnamefont {C.}~\bibnamefont
  {Boehm}}\ and\ \bibinfo {author} {\bibfnamefont {R.}~\bibnamefont
  {Schaeffer}},\ }\href {\doibase 10.1051/0004-6361:20042238} {\bibfield
  {journal} {\bibinfo  {journal} {Astron.Astrophys.}\ }\textbf {\bibinfo
  {volume} {438}},\ \bibinfo {pages} {419} (\bibinfo {year} {2005})},\ \Eprint
  {http://arxiv.org/abs/astro-ph/0410591} {arXiv:astro-ph/0410591 [astro-ph]}
  \BibitemShut {NoStop}%
\bibitem [{\citenamefont {Bertschinger}(2006)}]{Bertschinger:2006nq}%
  \BibitemOpen
  \bibfield  {author} {\bibinfo {author} {\bibfnamefont {E.}~\bibnamefont
  {Bertschinger}},\ }\href {\doibase 10.1103/PhysRevD.74.063509} {\bibfield
  {journal} {\bibinfo  {journal} {Phys.Rev.}\ }\textbf {\bibinfo {volume}
  {D74}},\ \bibinfo {pages} {063509} (\bibinfo {year} {2006})},\ \Eprint
  {http://arxiv.org/abs/astro-ph/0607319} {arXiv:astro-ph/0607319 [astro-ph]}
  \BibitemShut {NoStop}%
\bibitem [{\citenamefont {Mangano}\ \emph {et~al.}(2006)\citenamefont
  {Mangano}, \citenamefont {Melchiorri}, \citenamefont {Serra}, \citenamefont
  {Cooray},\ and\ \citenamefont {Kamionkowski}}]{Mangano:2006mp}%
  \BibitemOpen
  \bibfield  {author} {\bibinfo {author} {\bibfnamefont {G.}~\bibnamefont
  {Mangano}}, \bibinfo {author} {\bibfnamefont {A.}~\bibnamefont {Melchiorri}},
  \bibinfo {author} {\bibfnamefont {P.}~\bibnamefont {Serra}}, \bibinfo
  {author} {\bibfnamefont {A.}~\bibnamefont {Cooray}}, \ and\ \bibinfo {author}
  {\bibfnamefont {M.}~\bibnamefont {Kamionkowski}},\ }\href {\doibase
  10.1103/PhysRevD.74.043517} {\bibfield  {journal} {\bibinfo  {journal}
  {Phys.Rev.}\ }\textbf {\bibinfo {volume} {D74}},\ \bibinfo {pages} {043517}
  (\bibinfo {year} {2006})},\ \Eprint {http://arxiv.org/abs/astro-ph/0606190}
  {arXiv:astro-ph/0606190 [astro-ph]} \BibitemShut {NoStop}%
\bibitem [{\citenamefont {Serra}\ \emph {et~al.}(2010)\citenamefont {Serra},
  \citenamefont {Zalamea}, \citenamefont {Cooray}, \citenamefont {Mangano},\
  and\ \citenamefont {Melchiorri}}]{Serra:2009uu}%
  \BibitemOpen
  \bibfield  {author} {\bibinfo {author} {\bibfnamefont {P.}~\bibnamefont
  {Serra}}, \bibinfo {author} {\bibfnamefont {F.}~\bibnamefont {Zalamea}},
  \bibinfo {author} {\bibfnamefont {A.}~\bibnamefont {Cooray}}, \bibinfo
  {author} {\bibfnamefont {G.}~\bibnamefont {Mangano}}, \ and\ \bibinfo
  {author} {\bibfnamefont {A.}~\bibnamefont {Melchiorri}},\ }\href {\doibase
  10.1103/PhysRevD.81.043507} {\bibfield  {journal} {\bibinfo  {journal}
  {Phys.Rev.}\ }\textbf {\bibinfo {volume} {D81}},\ \bibinfo {pages} {043507}
  (\bibinfo {year} {2010})},\ \Eprint {http://arxiv.org/abs/0911.4411}
  {arXiv:0911.4411 [astro-ph.CO]} \BibitemShut {NoStop}%
\bibitem [{\citenamefont {Wilkinson}\ \emph
  {et~al.}(2014{\natexlab{a}})\citenamefont {Wilkinson}, \citenamefont
  {Boehm},\ and\ \citenamefont {Lesgourgues}}]{Wilkinson:2014ksa}%
  \BibitemOpen
  \bibfield  {author} {\bibinfo {author} {\bibfnamefont {R.~J.}\ \bibnamefont
  {Wilkinson}}, \bibinfo {author} {\bibfnamefont {C.}~\bibnamefont {Boehm}}, \
  and\ \bibinfo {author} {\bibfnamefont {J.}~\bibnamefont {Lesgourgues}},\
  }\href {\doibase 10.1088/1475-7516/2014/05/011} {\bibfield  {journal}
  {\bibinfo  {journal} {JCAP}\ }\textbf {\bibinfo {volume} {1405}},\ \bibinfo
  {pages} {011} (\bibinfo {year} {2014}{\natexlab{a}})},\ \Eprint
  {http://arxiv.org/abs/1401.7597} {arXiv:1401.7597 [astro-ph.CO]} \BibitemShut
  {NoStop}%
\bibitem [{\citenamefont {van~den Aarssen}\ \emph {et~al.}(2012)\citenamefont
  {van~den Aarssen}, \citenamefont {Bringmann},\ and\ \citenamefont
  {Pfrommer}}]{Aarssen:2012fx}%
  \BibitemOpen
  \bibfield  {author} {\bibinfo {author} {\bibfnamefont {L.~G.}\ \bibnamefont
  {van~den Aarssen}}, \bibinfo {author} {\bibfnamefont {T.}~\bibnamefont
  {Bringmann}}, \ and\ \bibinfo {author} {\bibfnamefont {C.}~\bibnamefont
  {Pfrommer}},\ }\href {\doibase 10.1103/PhysRevLett.109.231301} {\bibfield
  {journal} {\bibinfo  {journal} {Phys.Rev.Lett.}\ }\textbf {\bibinfo {volume}
  {109}},\ \bibinfo {pages} {231301} (\bibinfo {year} {2012})},\ \Eprint
  {http://arxiv.org/abs/1205.5809} {arXiv:1205.5809 [astro-ph.CO]} \BibitemShut
  {NoStop}%
\bibitem [{\citenamefont {Farzan}\ and\ \citenamefont
  {Palomares-Ruiz}(2014)}]{Farzan:2014gza}%
  \BibitemOpen
  \bibfield  {author} {\bibinfo {author} {\bibfnamefont {Y.}~\bibnamefont
  {Farzan}}\ and\ \bibinfo {author} {\bibfnamefont {S.}~\bibnamefont
  {Palomares-Ruiz}},\ }\href {\doibase 10.1088/1475-7516/2014/06/014}
  {\bibfield  {journal} {\bibinfo  {journal} {JCAP}\ }\textbf {\bibinfo
  {volume} {1406}},\ \bibinfo {pages} {014} (\bibinfo {year} {2014})},\ \Eprint
  {http://arxiv.org/abs/1401.7019} {arXiv:1401.7019 [hep-ph]} \BibitemShut
  {NoStop}%
\bibitem [{\citenamefont {Boehm}\ \emph {et~al.}(2014)\citenamefont {Boehm},
  \citenamefont {Schewtschenko}, \citenamefont {Wilkinson}, \citenamefont
  {Baugh},\ and\ \citenamefont {Pascoli}}]{Boehm:2014vja}%
  \BibitemOpen
  \bibfield  {author} {\bibinfo {author} {\bibfnamefont {C.}~\bibnamefont
  {Boehm}}, \bibinfo {author} {\bibfnamefont {J.}~\bibnamefont
  {Schewtschenko}}, \bibinfo {author} {\bibfnamefont {R.}~\bibnamefont
  {Wilkinson}}, \bibinfo {author} {\bibfnamefont {C.}~\bibnamefont {Baugh}}, \
  and\ \bibinfo {author} {\bibfnamefont {S.}~\bibnamefont {Pascoli}},\ }\href
  {\doibase 10.1093/mnrasl/slu115} {\bibfield  {journal} {\bibinfo  {journal}
  {Mon.Not.Roy.Astron.Soc.}\ }\textbf {\bibinfo {volume} {445}},\ \bibinfo
  {pages} {L31} (\bibinfo {year} {2014})},\ \Eprint
  {http://arxiv.org/abs/1404.7012} {arXiv:1404.7012 [astro-ph.CO]} \BibitemShut
  {NoStop}%
\bibitem [{\citenamefont {Cherry}\ \emph {et~al.}(2014)\citenamefont {Cherry},
  \citenamefont {Friedland},\ and\ \citenamefont {Shoemaker}}]{Cherry:2014xra}%
  \BibitemOpen
  \bibfield  {author} {\bibinfo {author} {\bibfnamefont {J.~F.}\ \bibnamefont
  {Cherry}}, \bibinfo {author} {\bibfnamefont {A.}~\bibnamefont {Friedland}}, \
  and\ \bibinfo {author} {\bibfnamefont {I.~M.}\ \bibnamefont {Shoemaker}},\
  }\href@noop {} {\  (\bibinfo {year} {2014})},\ \Eprint
  {http://arxiv.org/abs/1411.1071} {arXiv:1411.1071 [hep-ph]} \BibitemShut
  {NoStop}%
\bibitem [{\citenamefont {Bertoni}\ \emph {et~al.}(2015)\citenamefont
  {Bertoni}, \citenamefont {Ipek}, \citenamefont {McKeen},\ and\ \citenamefont
  {Nelson}}]{Bertoni:2014mva}%
  \BibitemOpen
  \bibfield  {author} {\bibinfo {author} {\bibfnamefont {B.}~\bibnamefont
  {Bertoni}}, \bibinfo {author} {\bibfnamefont {S.}~\bibnamefont {Ipek}},
  \bibinfo {author} {\bibfnamefont {D.}~\bibnamefont {McKeen}}, \ and\ \bibinfo
  {author} {\bibfnamefont {A.~E.}\ \bibnamefont {Nelson}},\ }\href {\doibase
  10.1007/JHEP04(2015)170} {\bibfield  {journal} {\bibinfo  {journal} {JHEP}\
  }\textbf {\bibinfo {volume} {1504}},\ \bibinfo {pages} {170} (\bibinfo {year}
  {2015})},\ \Eprint {http://arxiv.org/abs/1412.3113} {arXiv:1412.3113
  [hep-ph]} \BibitemShut {NoStop}%
\bibitem [{\citenamefont {Schewtschenko}\ \emph {et~al.}(2015)\citenamefont
  {Schewtschenko}, \citenamefont {Wilkinson}, \citenamefont {Baugh},
  \citenamefont {Boehm},\ and\ \citenamefont
  {Pascoli}}]{Schewtschenko:2014fca}%
  \BibitemOpen
  \bibfield  {author} {\bibinfo {author} {\bibfnamefont {J.}~\bibnamefont
  {Schewtschenko}}, \bibinfo {author} {\bibfnamefont {R.}~\bibnamefont
  {Wilkinson}}, \bibinfo {author} {\bibfnamefont {C.}~\bibnamefont {Baugh}},
  \bibinfo {author} {\bibfnamefont {C.}~\bibnamefont {Boehm}}, \ and\ \bibinfo
  {author} {\bibfnamefont {S.}~\bibnamefont {Pascoli}},\ }\href {\doibase
  10.1093/mnras/stv431} {\bibfield  {journal} {\bibinfo  {journal}
  {Mon.Not.Roy.Astron.Soc.}\ }\textbf {\bibinfo {volume} {449}},\ \bibinfo
  {pages} {3587} (\bibinfo {year} {2015})},\ \Eprint
  {http://arxiv.org/abs/1412.4905} {arXiv:1412.4905 [astro-ph.CO]} \BibitemShut
  {NoStop}%
\bibitem [{\citenamefont {Davis}\ and\ \citenamefont
  {Silk}(2015)}]{Davis:2015rza}%
  \BibitemOpen
  \bibfield  {author} {\bibinfo {author} {\bibfnamefont {J.~H.}\ \bibnamefont
  {Davis}}\ and\ \bibinfo {author} {\bibfnamefont {J.}~\bibnamefont {Silk}},\
  }\href@noop {} {\  (\bibinfo {year} {2015})},\ \Eprint
  {http://arxiv.org/abs/1505.01843} {arXiv:1505.01843 [hep-ph]} \BibitemShut
  {NoStop}%
\bibitem [{\citenamefont {Sigurdson}\ \emph {et~al.}(2004)\citenamefont
  {Sigurdson}, \citenamefont {Doran}, \citenamefont {Kurylov}, \citenamefont
  {Caldwell},\ and\ \citenamefont {Kamionkowski}}]{Sigurdson:2004zp}%
  \BibitemOpen
  \bibfield  {author} {\bibinfo {author} {\bibfnamefont {K.}~\bibnamefont
  {Sigurdson}}, \bibinfo {author} {\bibfnamefont {M.}~\bibnamefont {Doran}},
  \bibinfo {author} {\bibfnamefont {A.}~\bibnamefont {Kurylov}}, \bibinfo
  {author} {\bibfnamefont {R.~R.}\ \bibnamefont {Caldwell}}, \ and\ \bibinfo
  {author} {\bibfnamefont {M.}~\bibnamefont {Kamionkowski}},\ }\href {\doibase
  10.1103/PhysRevD.70.083501, 10.1103/PhysRevD.70.083501
  10.1103/PhysRevD.73.089903, 10.1103/PhysRevD.73.089903} {\bibfield  {journal}
  {\bibinfo  {journal} {Phys.Rev.}\ }\textbf {\bibinfo {volume} {D70}},\
  \bibinfo {pages} {083501} (\bibinfo {year} {2004})},\ \Eprint
  {http://arxiv.org/abs/astro-ph/0406355} {arXiv:astro-ph/0406355 [astro-ph]}
  \BibitemShut {NoStop}%
\bibitem [{\citenamefont {Wilkinson}\ \emph
  {et~al.}(2014{\natexlab{b}})\citenamefont {Wilkinson}, \citenamefont
  {Lesgourgues},\ and\ \citenamefont {Boehm}}]{Wilkinson:2013kia}%
  \BibitemOpen
  \bibfield  {author} {\bibinfo {author} {\bibfnamefont {R.~J.}\ \bibnamefont
  {Wilkinson}}, \bibinfo {author} {\bibfnamefont {J.}~\bibnamefont
  {Lesgourgues}}, \ and\ \bibinfo {author} {\bibfnamefont {C.}~\bibnamefont
  {Boehm}},\ }\href {\doibase 10.1088/1475-7516/2014/04/026} {\bibfield
  {journal} {\bibinfo  {journal} {JCAP}\ }\textbf {\bibinfo {volume} {1404}},\
  \bibinfo {pages} {026} (\bibinfo {year} {2014}{\natexlab{b}})},\ \Eprint
  {http://arxiv.org/abs/1309.7588} {arXiv:1309.7588 [astro-ph.CO]} \BibitemShut
  {NoStop}%
\bibitem [{\citenamefont {Dolgov}\ \emph {et~al.}(2013)\citenamefont {Dolgov},
  \citenamefont {Dubovsky}, \citenamefont {Rubtsov},\ and\ \citenamefont
  {Tkachev}}]{Dolgov:2013una}%
  \BibitemOpen
  \bibfield  {author} {\bibinfo {author} {\bibfnamefont {A.}~\bibnamefont
  {Dolgov}}, \bibinfo {author} {\bibfnamefont {S.}~\bibnamefont {Dubovsky}},
  \bibinfo {author} {\bibfnamefont {G.}~\bibnamefont {Rubtsov}}, \ and\
  \bibinfo {author} {\bibfnamefont {I.}~\bibnamefont {Tkachev}},\ }\href
  {\doibase 10.1103/PhysRevD.88.117701} {\bibfield  {journal} {\bibinfo
  {journal} {Phys.Rev.}\ }\textbf {\bibinfo {volume} {D88}},\ \bibinfo {pages}
  {117701} (\bibinfo {year} {2013})},\ \Eprint {http://arxiv.org/abs/1310.2376}
  {arXiv:1310.2376 [hep-ph]} \BibitemShut {NoStop}%
\bibitem [{\citenamefont {Chen}\ \emph {et~al.}(2002)\citenamefont {Chen},
  \citenamefont {Hannestad},\ and\ \citenamefont {Scherrer}}]{Chen:2002yh}%
  \BibitemOpen
  \bibfield  {author} {\bibinfo {author} {\bibfnamefont {X.-l.}\ \bibnamefont
  {Chen}}, \bibinfo {author} {\bibfnamefont {S.}~\bibnamefont {Hannestad}}, \
  and\ \bibinfo {author} {\bibfnamefont {R.~J.}\ \bibnamefont {Scherrer}},\
  }\href {\doibase 10.1103/PhysRevD.65.123515} {\bibfield  {journal} {\bibinfo
  {journal} {Phys.Rev.}\ }\textbf {\bibinfo {volume} {D65}},\ \bibinfo {pages}
  {123515} (\bibinfo {year} {2002})},\ \Eprint
  {http://arxiv.org/abs/astro-ph/0202496} {arXiv:astro-ph/0202496 [astro-ph]}
  \BibitemShut {NoStop}%
\bibitem [{\citenamefont {Dvorkin}\ \emph {et~al.}(2014)\citenamefont
  {Dvorkin}, \citenamefont {Blum},\ and\ \citenamefont
  {Kamionkowski}}]{Dvorkin:2013cea}%
  \BibitemOpen
  \bibfield  {author} {\bibinfo {author} {\bibfnamefont {C.}~\bibnamefont
  {Dvorkin}}, \bibinfo {author} {\bibfnamefont {K.}~\bibnamefont {Blum}}, \
  and\ \bibinfo {author} {\bibfnamefont {M.}~\bibnamefont {Kamionkowski}},\
  }\href {\doibase 10.1103/PhysRevD.89.023519} {\bibfield  {journal} {\bibinfo
  {journal} {Phys.Rev.}\ }\textbf {\bibinfo {volume} {D89}},\ \bibinfo {pages}
  {023519} (\bibinfo {year} {2014})},\ \Eprint {http://arxiv.org/abs/1311.2937}
  {arXiv:1311.2937 [astro-ph.CO]} \BibitemShut {NoStop}%
\bibitem [{\citenamefont {Park}\ \emph {et~al.}(2012)\citenamefont {Park},
  \citenamefont {Hwang},\ and\ \citenamefont {Noh}}]{Park:2012ru}%
  \BibitemOpen
  \bibfield  {author} {\bibinfo {author} {\bibfnamefont {C.-G.}\ \bibnamefont
  {Park}}, \bibinfo {author} {\bibfnamefont {J.-c.}\ \bibnamefont {Hwang}}, \
  and\ \bibinfo {author} {\bibfnamefont {H.}~\bibnamefont {Noh}},\ }\href
  {\doibase 10.1103/PhysRevD.86.083535} {\bibfield  {journal} {\bibinfo
  {journal} {Phys.Rev.}\ }\textbf {\bibinfo {volume} {D86}},\ \bibinfo {pages}
  {083535} (\bibinfo {year} {2012})},\ \Eprint {http://arxiv.org/abs/1207.3124}
  {arXiv:1207.3124 [astro-ph.CO]} \BibitemShut {NoStop}%
\bibitem [{\citenamefont {Diamanti}\ \emph {et~al.}(2013)\citenamefont
  {Diamanti}, \citenamefont {Giusarma}, \citenamefont {Mena}, \citenamefont
  {Archidiacono},\ and\ \citenamefont {Melchiorri}}]{Diamanti:2012tg}%
  \BibitemOpen
  \bibfield  {author} {\bibinfo {author} {\bibfnamefont {R.}~\bibnamefont
  {Diamanti}}, \bibinfo {author} {\bibfnamefont {E.}~\bibnamefont {Giusarma}},
  \bibinfo {author} {\bibfnamefont {O.}~\bibnamefont {Mena}}, \bibinfo {author}
  {\bibfnamefont {M.}~\bibnamefont {Archidiacono}}, \ and\ \bibinfo {author}
  {\bibfnamefont {A.}~\bibnamefont {Melchiorri}},\ }\href {\doibase
  10.1103/PhysRevD.87.063509} {\bibfield  {journal} {\bibinfo  {journal}
  {Phys.Rev.}\ }\textbf {\bibinfo {volume} {D87}},\ \bibinfo {pages} {063509}
  (\bibinfo {year} {2013})},\ \Eprint {http://arxiv.org/abs/1212.6007}
  {arXiv:1212.6007 [astro-ph.CO]} \BibitemShut {NoStop}%
\bibitem [{\citenamefont {Blennow}\ \emph {et~al.}(2012)\citenamefont
  {Blennow}, \citenamefont {Fernandez-Martinez}, \citenamefont {Mena},
  \citenamefont {Redondo},\ and\ \citenamefont {Serra}}]{Blennow:2012de}%
  \BibitemOpen
  \bibfield  {author} {\bibinfo {author} {\bibfnamefont {M.}~\bibnamefont
  {Blennow}}, \bibinfo {author} {\bibfnamefont {E.}~\bibnamefont
  {Fernandez-Martinez}}, \bibinfo {author} {\bibfnamefont {O.}~\bibnamefont
  {Mena}}, \bibinfo {author} {\bibfnamefont {J.}~\bibnamefont {Redondo}}, \
  and\ \bibinfo {author} {\bibfnamefont {P.}~\bibnamefont {Serra}},\ }\href
  {\doibase 10.1088/1475-7516/2012/07/022} {\bibfield  {journal} {\bibinfo
  {journal} {JCAP}\ }\textbf {\bibinfo {volume} {1207}},\ \bibinfo {pages}
  {022} (\bibinfo {year} {2012})},\ \Eprint {http://arxiv.org/abs/1203.5803}
  {arXiv:1203.5803 [hep-ph]} \BibitemShut {NoStop}%
\bibitem [{\citenamefont {Cyr-Racine}\ \emph {et~al.}(2014)\citenamefont
  {Cyr-Racine}, \citenamefont {de~Putter}, \citenamefont {Raccanelli},\ and\
  \citenamefont {Sigurdson}}]{Cyr-Racine:2013fsa}%
  \BibitemOpen
  \bibfield  {author} {\bibinfo {author} {\bibfnamefont {F.-Y.}\ \bibnamefont
  {Cyr-Racine}}, \bibinfo {author} {\bibfnamefont {R.}~\bibnamefont
  {de~Putter}}, \bibinfo {author} {\bibfnamefont {A.}~\bibnamefont
  {Raccanelli}}, \ and\ \bibinfo {author} {\bibfnamefont {K.}~\bibnamefont
  {Sigurdson}},\ }\href {\doibase 10.1103/PhysRevD.89.063517} {\bibfield
  {journal} {\bibinfo  {journal} {Phys.Rev.}\ }\textbf {\bibinfo {volume}
  {D89}},\ \bibinfo {pages} {063517} (\bibinfo {year} {2014})},\ \Eprint
  {http://arxiv.org/abs/1310.3278} {arXiv:1310.3278 [astro-ph.CO]} \BibitemShut
  {NoStop}%
\bibitem [{\citenamefont {Bouchet}\ \emph {et~al.}(2011)\citenamefont {Bouchet}
  \emph {et~al.}}]{Bouchet:2011ck}%
  \BibitemOpen
  \bibfield  {author} {\bibinfo {author} {\bibfnamefont {F.}~\bibnamefont
  {Bouchet}} \emph {et~al.} (\bibinfo {collaboration} {COrE Collaboration}),\
  }\href@noop {} {\  (\bibinfo {year} {2011})},\ \Eprint
  {http://arxiv.org/abs/1102.2181} {arXiv:1102.2181 [astro-ph.CO]} \BibitemShut
  {NoStop}%
\bibitem [{\citenamefont {Kogut}\ \emph {et~al.}(2011)\citenamefont {Kogut},
  \citenamefont {Fixsen}, \citenamefont {Chuss}, \citenamefont {Dotson},
  \citenamefont {Dwek} \emph {et~al.}}]{Kogut:2011xw}%
  \BibitemOpen
  \bibfield  {author} {\bibinfo {author} {\bibfnamefont {A.}~\bibnamefont
  {Kogut}}, \bibinfo {author} {\bibfnamefont {D.}~\bibnamefont {Fixsen}},
  \bibinfo {author} {\bibfnamefont {D.}~\bibnamefont {Chuss}}, \bibinfo
  {author} {\bibfnamefont {J.}~\bibnamefont {Dotson}}, \bibinfo {author}
  {\bibfnamefont {E.}~\bibnamefont {Dwek}},  \emph {et~al.},\ }\href {\doibase
  10.1088/1475-7516/2011/07/025} {\bibfield  {journal} {\bibinfo  {journal}
  {JCAP}\ }\textbf {\bibinfo {volume} {1107}},\ \bibinfo {pages} {025}
  (\bibinfo {year} {2011})},\ \Eprint {http://arxiv.org/abs/1105.2044}
  {arXiv:1105.2044 [astro-ph.CO]} \BibitemShut {NoStop}%
\bibitem [{\citenamefont {Anderson}\ \emph
  {et~al.}(2014{\natexlab{a}})\citenamefont {Anderson} \emph
  {et~al.}}]{Anderson:2013zyy}%
  \BibitemOpen
  \bibfield  {author} {\bibinfo {author} {\bibfnamefont {L.}~\bibnamefont
  {Anderson}} \emph {et~al.} (\bibinfo {collaboration} {BOSS}),\ }\href
  {\doibase 10.1093/mnras/stu523} {\bibfield  {journal} {\bibinfo  {journal}
  {Mon.Not.Roy.Astron.Soc.}\ }\textbf {\bibinfo {volume} {441}},\ \bibinfo
  {pages} {24} (\bibinfo {year} {2014}{\natexlab{a}})},\ \Eprint
  {http://arxiv.org/abs/1312.4877} {arXiv:1312.4877 [astro-ph.CO]} \BibitemShut
  {NoStop}%
\bibitem [{\citenamefont {Anderson}\ \emph
  {et~al.}(2014{\natexlab{b}})\citenamefont {Anderson}, \citenamefont
  {Aubourg}, \citenamefont {Bailey}, \citenamefont {Beutler}, \citenamefont
  {Bolton} \emph {et~al.}}]{Anderson21122012}%
  \BibitemOpen
  \bibfield  {author} {\bibinfo {author} {\bibfnamefont {L.}~\bibnamefont
  {Anderson}}, \bibinfo {author} {\bibfnamefont {E.}~\bibnamefont {Aubourg}},
  \bibinfo {author} {\bibfnamefont {S.}~\bibnamefont {Bailey}}, \bibinfo
  {author} {\bibfnamefont {F.}~\bibnamefont {Beutler}}, \bibinfo {author}
  {\bibfnamefont {A.~S.}\ \bibnamefont {Bolton}},  \emph {et~al.},\ }\href
  {\doibase 10.1093/mnras/stt2206} {\bibfield  {journal} {\bibinfo  {journal}
  {Mon.Not.Roy.Astron.Soc.}\ }\textbf {\bibinfo {volume} {439}},\ \bibinfo
  {pages} {83} (\bibinfo {year} {2014}{\natexlab{b}})},\ \Eprint
  {http://arxiv.org/abs/1303.4666} {arXiv:1303.4666 [astro-ph.CO]} \BibitemShut
  {NoStop}%
\bibitem [{\citenamefont {Beutler}\ \emph {et~al.}(2011)\citenamefont
  {Beutler}, \citenamefont {Blake}, \citenamefont {Colless}, \citenamefont
  {Jones}, \citenamefont {Staveley-Smith} \emph {et~al.}}]{Beutler01102011}%
  \BibitemOpen
  \bibfield  {author} {\bibinfo {author} {\bibfnamefont {F.}~\bibnamefont
  {Beutler}}, \bibinfo {author} {\bibfnamefont {C.}~\bibnamefont {Blake}},
  \bibinfo {author} {\bibfnamefont {M.}~\bibnamefont {Colless}}, \bibinfo
  {author} {\bibfnamefont {D.~H.}\ \bibnamefont {Jones}}, \bibinfo {author}
  {\bibfnamefont {L.}~\bibnamefont {Staveley-Smith}},  \emph {et~al.},\ }\href
  {\doibase 10.1111/j.1365-2966.2011.19250.x} {\bibfield  {journal} {\bibinfo
  {journal} {Mon.Not.Roy.Astron.Soc.}\ }\textbf {\bibinfo {volume} {416}},\
  \bibinfo {pages} {3017} (\bibinfo {year} {2011})},\ \Eprint
  {http://arxiv.org/abs/1106.3366} {arXiv:1106.3366 [astro-ph.CO]} \BibitemShut
  {NoStop}%
\bibitem [{\citenamefont {Blake}\ \emph {et~al.}(2011)\citenamefont {Blake},
  \citenamefont {Kazin}, \citenamefont {Beutler}, \citenamefont {Davis},
  \citenamefont {Parkinson} \emph {et~al.}}]{Blake11122011}%
  \BibitemOpen
  \bibfield  {author} {\bibinfo {author} {\bibfnamefont {C.}~\bibnamefont
  {Blake}}, \bibinfo {author} {\bibfnamefont {E.}~\bibnamefont {Kazin}},
  \bibinfo {author} {\bibfnamefont {F.}~\bibnamefont {Beutler}}, \bibinfo
  {author} {\bibfnamefont {T.}~\bibnamefont {Davis}}, \bibinfo {author}
  {\bibfnamefont {D.}~\bibnamefont {Parkinson}},  \emph {et~al.},\ }\href
  {\doibase 10.1111/j.1365-2966.2011.19592.x} {\bibfield  {journal} {\bibinfo
  {journal} {Mon.Not.Roy.Astron.Soc.}\ }\textbf {\bibinfo {volume} {418}},\
  \bibinfo {pages} {1707} (\bibinfo {year} {2011})},\ \Eprint
  {http://arxiv.org/abs/1108.2635} {arXiv:1108.2635 [astro-ph.CO]} \BibitemShut
  {NoStop}%
\bibitem [{\citenamefont {Padmanabhan}\ \emph {et~al.}(2012)\citenamefont
  {Padmanabhan}, \citenamefont {Xu}, \citenamefont {Eisenstein}, \citenamefont
  {Scalzo}, \citenamefont {Cuesta} \emph {et~al.}}]{Padmanabhan11122012}%
  \BibitemOpen
  \bibfield  {author} {\bibinfo {author} {\bibfnamefont {N.}~\bibnamefont
  {Padmanabhan}}, \bibinfo {author} {\bibfnamefont {X.}~\bibnamefont {Xu}},
  \bibinfo {author} {\bibfnamefont {D.~J.}\ \bibnamefont {Eisenstein}},
  \bibinfo {author} {\bibfnamefont {R.}~\bibnamefont {Scalzo}}, \bibinfo
  {author} {\bibfnamefont {A.~J.}\ \bibnamefont {Cuesta}},  \emph {et~al.},\
  }\href {\doibase 10.1111/j.1365-2966.2012.21888.x} {\bibfield  {journal}
  {\bibinfo  {journal} {Mon.Not.Roy.Astron.Soc.}\ }\textbf {\bibinfo {volume}
  {427}},\ \bibinfo {pages} {2132} (\bibinfo {year} {2012})},\ \Eprint
  {http://arxiv.org/abs/1202.0090} {arXiv:1202.0090 [astro-ph.CO]} \BibitemShut
  {NoStop}%
\bibitem [{\citenamefont {Percival}\ \emph {et~al.}(2010)\citenamefont
  {Percival} \emph {et~al.}}]{Percival01022010}%
  \BibitemOpen
  \bibfield  {author} {\bibinfo {author} {\bibfnamefont {W.~J.}\ \bibnamefont
  {Percival}} \emph {et~al.} (\bibinfo {collaboration} {SDSS}),\ }\href
  {\doibase 10.1111/j.1365-2966.2009.15812.x} {\bibfield  {journal} {\bibinfo
  {journal} {Mon.Not.Roy.Astron.Soc.}\ }\textbf {\bibinfo {volume} {401}},\
  \bibinfo {pages} {2148} (\bibinfo {year} {2010})},\ \Eprint
  {http://arxiv.org/abs/0907.1660} {arXiv:0907.1660 [astro-ph.CO]} \BibitemShut
  {NoStop}%
\bibitem [{\citenamefont {Parkinson}\ \emph {et~al.}(2012)\citenamefont
  {Parkinson}, \citenamefont {Riemer-Sorensen}, \citenamefont {Blake},
  \citenamefont {Poole}, \citenamefont {Davis} \emph
  {et~al.}}]{Parkinson:2012vd}%
  \BibitemOpen
  \bibfield  {author} {\bibinfo {author} {\bibfnamefont {D.}~\bibnamefont
  {Parkinson}}, \bibinfo {author} {\bibfnamefont {S.}~\bibnamefont
  {Riemer-Sorensen}}, \bibinfo {author} {\bibfnamefont {C.}~\bibnamefont
  {Blake}}, \bibinfo {author} {\bibfnamefont {G.~B.}\ \bibnamefont {Poole}},
  \bibinfo {author} {\bibfnamefont {T.~M.}\ \bibnamefont {Davis}},  \emph
  {et~al.},\ }\href {\doibase 10.1103/PhysRevD.86.103518} {\bibfield  {journal}
  {\bibinfo  {journal} {Phys.Rev.}\ }\textbf {\bibinfo {volume} {D86}},\
  \bibinfo {pages} {103518} (\bibinfo {year} {2012})},\ \Eprint
  {http://arxiv.org/abs/1210.2130} {arXiv:1210.2130 [astro-ph.CO]} \BibitemShut
  {NoStop}%
\bibitem [{\citenamefont {Dawson}\ \emph {et~al.}(2013)\citenamefont {Dawson}
  \emph {et~al.}}]{Dawson:2012va}%
  \BibitemOpen
  \bibfield  {author} {\bibinfo {author} {\bibfnamefont {K.~S.}\ \bibnamefont
  {Dawson}} \emph {et~al.} (\bibinfo {collaboration} {BOSS}),\ }\href {\doibase
  10.1088/0004-6256/145/1/10} {\bibfield  {journal} {\bibinfo  {journal}
  {Astron.J.}\ }\textbf {\bibinfo {volume} {145}},\ \bibinfo {pages} {10}
  (\bibinfo {year} {2013})},\ \Eprint {http://arxiv.org/abs/1208.0022}
  {arXiv:1208.0022 [astro-ph.CO]} \BibitemShut {NoStop}%
\bibitem [{\citenamefont {Hamann}\ \emph {et~al.}(2010)\citenamefont {Hamann},
  \citenamefont {Hannestad}, \citenamefont {Lesgourgues}, \citenamefont
  {Rampf},\ and\ \citenamefont {Wong}}]{Hamann:2010pw}%
  \BibitemOpen
  \bibfield  {author} {\bibinfo {author} {\bibfnamefont {J.}~\bibnamefont
  {Hamann}}, \bibinfo {author} {\bibfnamefont {S.}~\bibnamefont {Hannestad}},
  \bibinfo {author} {\bibfnamefont {J.}~\bibnamefont {Lesgourgues}}, \bibinfo
  {author} {\bibfnamefont {C.}~\bibnamefont {Rampf}}, \ and\ \bibinfo {author}
  {\bibfnamefont {Y.~Y.}\ \bibnamefont {Wong}},\ }\href {\doibase
  10.1088/1475-7516/2010/07/022} {\bibfield  {journal} {\bibinfo  {journal}
  {JCAP}\ }\textbf {\bibinfo {volume} {1007}},\ \bibinfo {pages} {022}
  (\bibinfo {year} {2010})},\ \Eprint {http://arxiv.org/abs/1003.3999}
  {arXiv:1003.3999 [astro-ph.CO]} \BibitemShut {NoStop}%
\bibitem [{\citenamefont {Giusarma}\ \emph
  {et~al.}(2013{\natexlab{a}})\citenamefont {Giusarma}, \citenamefont
  {De~Putter},\ and\ \citenamefont {Mena}}]{Giusarma:2012ph}%
  \BibitemOpen
  \bibfield  {author} {\bibinfo {author} {\bibfnamefont {E.}~\bibnamefont
  {Giusarma}}, \bibinfo {author} {\bibfnamefont {R.}~\bibnamefont {De~Putter}},
  \ and\ \bibinfo {author} {\bibfnamefont {O.}~\bibnamefont {Mena}},\ }\href
  {\doibase 10.1103/PhysRevD.87.043515} {\bibfield  {journal} {\bibinfo
  {journal} {Phys.Rev.}\ }\textbf {\bibinfo {volume} {D87}},\ \bibinfo {pages}
  {043515} (\bibinfo {year} {2013}{\natexlab{a}})},\ \Eprint
  {http://arxiv.org/abs/1211.2154} {arXiv:1211.2154 [astro-ph.CO]} \BibitemShut
  {NoStop}%
\bibitem [{\citenamefont {Giusarma}\ \emph
  {et~al.}(2013{\natexlab{b}})\citenamefont {Giusarma}, \citenamefont
  {de~Putter}, \citenamefont {Ho},\ and\ \citenamefont
  {Mena}}]{Giusarma:2013pmn}%
  \BibitemOpen
  \bibfield  {author} {\bibinfo {author} {\bibfnamefont {E.}~\bibnamefont
  {Giusarma}}, \bibinfo {author} {\bibfnamefont {R.}~\bibnamefont {de~Putter}},
  \bibinfo {author} {\bibfnamefont {S.}~\bibnamefont {Ho}}, \ and\ \bibinfo
  {author} {\bibfnamefont {O.}~\bibnamefont {Mena}},\ }\href {\doibase
  10.1103/PhysRevD.88.063515} {\bibfield  {journal} {\bibinfo  {journal}
  {Phys.Rev.}\ }\textbf {\bibinfo {volume} {D88}},\ \bibinfo {pages} {063515}
  (\bibinfo {year} {2013}{\natexlab{b}})},\ \Eprint
  {http://arxiv.org/abs/1306.5544} {arXiv:1306.5544 [astro-ph.CO]} \BibitemShut
  {NoStop}%
\bibitem [{\citenamefont {Levi}\ \emph {et~al.}(2013)\citenamefont {Levi} \emph
  {et~al.}}]{Levi:2013gra}%
  \BibitemOpen
  \bibfield  {author} {\bibinfo {author} {\bibfnamefont {M.}~\bibnamefont
  {Levi}} \emph {et~al.} (\bibinfo {collaboration} {DESI}),\ }\href@noop {} {\
  (\bibinfo {year} {2013})},\ \Eprint {http://arxiv.org/abs/1308.0847}
  {arXiv:1308.0847 [astro-ph.CO]} \BibitemShut {NoStop}%
\bibitem [{\citenamefont {Ma}\ and\ \citenamefont
  {Bertschinger}(1995)}]{Ma:1995ey}%
  \BibitemOpen
  \bibfield  {author} {\bibinfo {author} {\bibfnamefont {C.-P.}\ \bibnamefont
  {Ma}}\ and\ \bibinfo {author} {\bibfnamefont {E.}~\bibnamefont
  {Bertschinger}},\ }\href {\doibase 10.1086/176550} {\bibfield  {journal}
  {\bibinfo  {journal} {Astrophys.J.}\ }\textbf {\bibinfo {volume} {455}},\
  \bibinfo {pages} {7} (\bibinfo {year} {1995})},\ \Eprint
  {http://arxiv.org/abs/astro-ph/9506072} {arXiv:astro-ph/9506072 [astro-ph]}
  \BibitemShut {NoStop}%
\bibitem [{\citenamefont {Atrio-Barandela}\ and\ \citenamefont
  {Davidson}(1997)}]{AtrioBarandela:1996ur}%
  \BibitemOpen
  \bibfield  {author} {\bibinfo {author} {\bibfnamefont {F.}~\bibnamefont
  {Atrio-Barandela}}\ and\ \bibinfo {author} {\bibfnamefont {S.}~\bibnamefont
  {Davidson}},\ }\href {\doibase 10.1103/PhysRevD.55.5886} {\bibfield
  {journal} {\bibinfo  {journal} {Phys. Rev.}\ }\textbf {\bibinfo {volume}
  {D55}},\ \bibinfo {pages} {5886} (\bibinfo {year} {1997})},\ \Eprint
  {http://arxiv.org/abs/astro-ph/9702236} {arXiv:astro-ph/9702236 [astro-ph]}
  \BibitemShut {NoStop}%
\bibitem [{\citenamefont {Das}\ and\ \citenamefont
  {Weiner}(2011)}]{Das:2006ht}%
  \BibitemOpen
  \bibfield  {author} {\bibinfo {author} {\bibfnamefont {S.}~\bibnamefont
  {Das}}\ and\ \bibinfo {author} {\bibfnamefont {N.}~\bibnamefont {Weiner}},\
  }\href {\doibase 10.1103/PhysRevD.84.123511} {\bibfield  {journal} {\bibinfo
  {journal} {Phys. Rev.}\ }\textbf {\bibinfo {volume} {D84}},\ \bibinfo {pages}
  {123511} (\bibinfo {year} {2011})},\ \Eprint
  {http://arxiv.org/abs/astro-ph/0611353} {arXiv:astro-ph/0611353 [astro-ph]}
  \BibitemShut {NoStop}%
\bibitem [{\citenamefont {Cyr-Racine}\ and\ \citenamefont
  {Sigurdson}(2013)}]{CyrRacine:2012fz}%
  \BibitemOpen
  \bibfield  {author} {\bibinfo {author} {\bibfnamefont {F.-Y.}\ \bibnamefont
  {Cyr-Racine}}\ and\ \bibinfo {author} {\bibfnamefont {K.}~\bibnamefont
  {Sigurdson}},\ }\href {\doibase 10.1103/PhysRevD.87.103515} {\bibfield
  {journal} {\bibinfo  {journal} {Phys.Rev.}\ }\textbf {\bibinfo {volume}
  {D87}},\ \bibinfo {pages} {103515} (\bibinfo {year} {2013})},\ \Eprint
  {http://arxiv.org/abs/1209.5752} {arXiv:1209.5752 [astro-ph.CO]} \BibitemShut
  {NoStop}%
\bibitem [{\citenamefont {Buckley}\ \emph {et~al.}(2014)\citenamefont
  {Buckley}, \citenamefont {Zavala}, \citenamefont {Cyr-Racine}, \citenamefont
  {Sigurdson},\ and\ \citenamefont {Vogelsberger}}]{Buckley:2014hja}%
  \BibitemOpen
  \bibfield  {author} {\bibinfo {author} {\bibfnamefont {M.~R.}\ \bibnamefont
  {Buckley}}, \bibinfo {author} {\bibfnamefont {J.}~\bibnamefont {Zavala}},
  \bibinfo {author} {\bibfnamefont {F.-Y.}\ \bibnamefont {Cyr-Racine}},
  \bibinfo {author} {\bibfnamefont {K.}~\bibnamefont {Sigurdson}}, \ and\
  \bibinfo {author} {\bibfnamefont {M.}~\bibnamefont {Vogelsberger}},\ }\href
  {\doibase 10.1103/PhysRevD.90.043524} {\bibfield  {journal} {\bibinfo
  {journal} {Phys. Rev.}\ }\textbf {\bibinfo {volume} {D90}},\ \bibinfo {pages}
  {043524} (\bibinfo {year} {2014})},\ \Eprint {http://arxiv.org/abs/1405.2075}
  {arXiv:1405.2075 [astro-ph.CO]} \BibitemShut {NoStop}%
\bibitem [{\citenamefont {Viel}\ \emph {et~al.}(2005)\citenamefont {Viel},
  \citenamefont {Lesgourgues}, \citenamefont {Haehnelt}, \citenamefont
  {Matarrese},\ and\ \citenamefont {Riotto}}]{Viel:2005qj}%
  \BibitemOpen
  \bibfield  {author} {\bibinfo {author} {\bibfnamefont {M.}~\bibnamefont
  {Viel}}, \bibinfo {author} {\bibfnamefont {J.}~\bibnamefont {Lesgourgues}},
  \bibinfo {author} {\bibfnamefont {M.~G.}\ \bibnamefont {Haehnelt}}, \bibinfo
  {author} {\bibfnamefont {S.}~\bibnamefont {Matarrese}}, \ and\ \bibinfo
  {author} {\bibfnamefont {A.}~\bibnamefont {Riotto}},\ }\href {\doibase
  10.1103/PhysRevD.71.063534} {\bibfield  {journal} {\bibinfo  {journal} {Phys.
  Rev.}\ }\textbf {\bibinfo {volume} {D71}},\ \bibinfo {pages} {063534}
  (\bibinfo {year} {2005})},\ \Eprint {http://arxiv.org/abs/astro-ph/0501562}
  {arXiv:astro-ph/0501562 [astro-ph]} \BibitemShut {NoStop}%
\bibitem [{\citenamefont {Hlozek}\ \emph {et~al.}(2015)\citenamefont {Hlozek},
  \citenamefont {Grin}, \citenamefont {Marsh},\ and\ \citenamefont
  {Ferreira}}]{Hlozek:2014lca}%
  \BibitemOpen
  \bibfield  {author} {\bibinfo {author} {\bibfnamefont {R.}~\bibnamefont
  {Hlozek}}, \bibinfo {author} {\bibfnamefont {D.}~\bibnamefont {Grin}},
  \bibinfo {author} {\bibfnamefont {D.~J.}\ \bibnamefont {Marsh}}, \ and\
  \bibinfo {author} {\bibfnamefont {P.~G.}\ \bibnamefont {Ferreira}},\ }\href
  {\doibase 10.1103/PhysRevD.91.103512} {\bibfield  {journal} {\bibinfo
  {journal} {Phys. Rev.}\ }\textbf {\bibinfo {volume} {D91}},\ \bibinfo {pages}
  {103512} (\bibinfo {year} {2015})},\ \Eprint {http://arxiv.org/abs/1410.2896}
  {arXiv:1410.2896 [astro-ph.CO]} \BibitemShut {NoStop}%
\bibitem [{\citenamefont {Lesgourgues}(2011)}]{Lesgourgues:2011re}%
  \BibitemOpen
  \bibfield  {author} {\bibinfo {author} {\bibfnamefont {J.}~\bibnamefont
  {Lesgourgues}},\ }\href@noop {} {\  (\bibinfo {year} {2011})},\ \Eprint
  {http://arxiv.org/abs/1104.2932} {arXiv:1104.2932 [astro-ph.IM]} \BibitemShut
  {NoStop}%
\bibitem [{\citenamefont {Audren}\ \emph {et~al.}(2013)\citenamefont {Audren},
  \citenamefont {Lesgourgues}, \citenamefont {Benabed},\ and\ \citenamefont
  {Prunet}}]{Audren:2012wb}%
  \BibitemOpen
  \bibfield  {author} {\bibinfo {author} {\bibfnamefont {B.}~\bibnamefont
  {Audren}}, \bibinfo {author} {\bibfnamefont {J.}~\bibnamefont {Lesgourgues}},
  \bibinfo {author} {\bibfnamefont {K.}~\bibnamefont {Benabed}}, \ and\
  \bibinfo {author} {\bibfnamefont {S.}~\bibnamefont {Prunet}},\ }\href
  {\doibase 10.1088/1475-7516/2013/02/001} {\bibfield  {journal} {\bibinfo
  {journal} {JCAP}\ }\textbf {\bibinfo {volume} {1302}},\ \bibinfo {pages}
  {001} (\bibinfo {year} {2013})},\ \Eprint {http://arxiv.org/abs/1210.7183}
  {arXiv:1210.7183 [astro-ph.CO]} \BibitemShut {NoStop}%
\bibitem [{\citenamefont {Mangano}\ \emph {et~al.}(2005)\citenamefont
  {Mangano}, \citenamefont {Miele}, \citenamefont {Pastor}, \citenamefont
  {Pinto}, \citenamefont {Pisanti} \emph {et~al.}}]{Mangano:2005cc}%
  \BibitemOpen
  \bibfield  {author} {\bibinfo {author} {\bibfnamefont {G.}~\bibnamefont
  {Mangano}}, \bibinfo {author} {\bibfnamefont {G.}~\bibnamefont {Miele}},
  \bibinfo {author} {\bibfnamefont {S.}~\bibnamefont {Pastor}}, \bibinfo
  {author} {\bibfnamefont {T.}~\bibnamefont {Pinto}}, \bibinfo {author}
  {\bibfnamefont {O.}~\bibnamefont {Pisanti}},  \emph {et~al.},\ }\href
  {\doibase 10.1016/j.nuclphysb.2005.09.041} {\bibfield  {journal} {\bibinfo
  {journal} {Nucl.Phys.}\ }\textbf {\bibinfo {volume} {B729}},\ \bibinfo
  {pages} {221} (\bibinfo {year} {2005})},\ \Eprint
  {http://arxiv.org/abs/hep-ph/0506164} {arXiv:hep-ph/0506164 [hep-ph]}
  \BibitemShut {NoStop}%
\bibitem [{\citenamefont {Reid}\ \emph {et~al.}(2010)\citenamefont {Reid},
  \citenamefont {Percival}, \citenamefont {Eisenstein}, \citenamefont {Verde},
  \citenamefont {Spergel} \emph {et~al.}}]{Reid:2009xm}%
  \BibitemOpen
  \bibfield  {author} {\bibinfo {author} {\bibfnamefont {B.~A.}\ \bibnamefont
  {Reid}}, \bibinfo {author} {\bibfnamefont {W.~J.}\ \bibnamefont {Percival}},
  \bibinfo {author} {\bibfnamefont {D.~J.}\ \bibnamefont {Eisenstein}},
  \bibinfo {author} {\bibfnamefont {L.}~\bibnamefont {Verde}}, \bibinfo
  {author} {\bibfnamefont {D.~N.}\ \bibnamefont {Spergel}},  \emph {et~al.},\
  }\href {\doibase 10.1111/j.1365-2966.2010.16276.x} {\bibfield  {journal}
  {\bibinfo  {journal} {Mon.Not.Roy.Astron.Soc.}\ }\textbf {\bibinfo {volume}
  {404}},\ \bibinfo {pages} {60} (\bibinfo {year} {2010})},\ \Eprint
  {http://arxiv.org/abs/0907.1659} {arXiv:0907.1659 [astro-ph.CO]} \BibitemShut
  {NoStop}%
\bibitem [{\citenamefont {Riemer-Sorensen}\ \emph {et~al.}(2012)\citenamefont
  {Riemer-Sorensen}, \citenamefont {Blake}, \citenamefont {Parkinson},
  \citenamefont {Davis}, \citenamefont {Brough} \emph
  {et~al.}}]{RiemerSorensen:2011fe}%
  \BibitemOpen
  \bibfield  {author} {\bibinfo {author} {\bibfnamefont {S.}~\bibnamefont
  {Riemer-Sorensen}}, \bibinfo {author} {\bibfnamefont {C.}~\bibnamefont
  {Blake}}, \bibinfo {author} {\bibfnamefont {D.}~\bibnamefont {Parkinson}},
  \bibinfo {author} {\bibfnamefont {T.~M.}\ \bibnamefont {Davis}}, \bibinfo
  {author} {\bibfnamefont {S.}~\bibnamefont {Brough}},  \emph {et~al.},\ }\href
  {\doibase 10.1103/PhysRevD.85.081101} {\bibfield  {journal} {\bibinfo
  {journal} {Phys.Rev.}\ }\textbf {\bibinfo {volume} {D85}},\ \bibinfo {pages}
  {081101} (\bibinfo {year} {2012})},\ \Eprint {http://arxiv.org/abs/1112.4940}
  {arXiv:1112.4940 [astro-ph.CO]} \BibitemShut {NoStop}%
\bibitem [{\citenamefont {Tegmark}\ \emph {et~al.}(2006)\citenamefont {Tegmark}
  \emph {et~al.}}]{Tegmark:2006az}%
  \BibitemOpen
  \bibfield  {author} {\bibinfo {author} {\bibfnamefont {M.}~\bibnamefont
  {Tegmark}} \emph {et~al.} (\bibinfo {collaboration} {SDSS}),\ }\href
  {\doibase 10.1103/PhysRevD.74.123507} {\bibfield  {journal} {\bibinfo
  {journal} {Phys.Rev.}\ }\textbf {\bibinfo {volume} {D74}},\ \bibinfo {pages}
  {123507} (\bibinfo {year} {2006})},\ \Eprint
  {http://arxiv.org/abs/astro-ph/0608632} {arXiv:astro-ph/0608632 [astro-ph]}
  \BibitemShut {NoStop}%
\bibitem [{\citenamefont {Lewis}\ and\ \citenamefont
  {Bridle}(2002)}]{Lewis:2002ah}%
  \BibitemOpen
  \bibfield  {author} {\bibinfo {author} {\bibfnamefont {A.}~\bibnamefont
  {Lewis}}\ and\ \bibinfo {author} {\bibfnamefont {S.}~\bibnamefont {Bridle}},\
  }\href {\doibase 10.1103/PhysRevD.66.103511} {\bibfield  {journal} {\bibinfo
  {journal} {Phys.Rev.}\ }\textbf {\bibinfo {volume} {D66}},\ \bibinfo {pages}
  {103511} (\bibinfo {year} {2002})},\ \Eprint
  {http://arxiv.org/abs/astro-ph/0205436} {arXiv:astro-ph/0205436 [astro-ph]}
  \BibitemShut {NoStop}%
\bibitem [{\citenamefont {Perotto}\ \emph {et~al.}(2006)\citenamefont
  {Perotto}, \citenamefont {Lesgourgues}, \citenamefont {Hannestad},
  \citenamefont {Tu},\ and\ \citenamefont {Wong}}]{Perotto:2006rj}%
  \BibitemOpen
  \bibfield  {author} {\bibinfo {author} {\bibfnamefont {L.}~\bibnamefont
  {Perotto}}, \bibinfo {author} {\bibfnamefont {J.}~\bibnamefont
  {Lesgourgues}}, \bibinfo {author} {\bibfnamefont {S.}~\bibnamefont
  {Hannestad}}, \bibinfo {author} {\bibfnamefont {H.}~\bibnamefont {Tu}}, \
  and\ \bibinfo {author} {\bibfnamefont {Y.~Y.}\ \bibnamefont {Wong}},\ }\href
  {\doibase 10.1088/1475-7516/2006/10/013} {\bibfield  {journal} {\bibinfo
  {journal} {JCAP}\ }\textbf {\bibinfo {volume} {0610}},\ \bibinfo {pages}
  {013} (\bibinfo {year} {2006})},\ \Eprint
  {http://arxiv.org/abs/astro-ph/0606227} {arXiv:astro-ph/0606227 [astro-ph]}
  \BibitemShut {NoStop}%
\bibitem [{\citenamefont {Ade}\ \emph {et~al.}(2015{\natexlab{a}})\citenamefont
  {Ade} \emph {et~al.}}]{Ade:2015lrj}%
  \BibitemOpen
  \bibfield  {author} {\bibinfo {author} {\bibfnamefont {P.}~\bibnamefont
  {Ade}} \emph {et~al.} (\bibinfo {collaboration} {Planck}),\ }\href@noop {} {\
   (\bibinfo {year} {2015}{\natexlab{a}})},\ \Eprint
  {http://arxiv.org/abs/1502.02114} {arXiv:1502.02114 [astro-ph.CO]}
  \BibitemShut {NoStop}%
\bibitem [{\citenamefont {Tegmark}\ \emph {et~al.}(1997)\citenamefont
  {Tegmark}, \citenamefont {Taylor},\ and\ \citenamefont
  {Heavens}}]{Tegmark:1996bz}%
  \BibitemOpen
  \bibfield  {author} {\bibinfo {author} {\bibfnamefont {M.}~\bibnamefont
  {Tegmark}}, \bibinfo {author} {\bibfnamefont {A.}~\bibnamefont {Taylor}}, \
  and\ \bibinfo {author} {\bibfnamefont {A.}~\bibnamefont {Heavens}},\ }\href
  {\doibase 10.1086/303939} {\bibfield  {journal} {\bibinfo  {journal}
  {Astrophys.J.}\ }\textbf {\bibinfo {volume} {480}},\ \bibinfo {pages} {22}
  (\bibinfo {year} {1997})},\ \Eprint {http://arxiv.org/abs/astro-ph/9603021}
  {arXiv:astro-ph/9603021 [astro-ph]} \BibitemShut {NoStop}%
\bibitem [{\citenamefont {Jungman}\ \emph {et~al.}(1996)\citenamefont
  {Jungman}, \citenamefont {Kamionkowski}, \citenamefont {Kosowsky},\ and\
  \citenamefont {Spergel}}]{Jungman:1995bz}%
  \BibitemOpen
  \bibfield  {author} {\bibinfo {author} {\bibfnamefont {G.}~\bibnamefont
  {Jungman}}, \bibinfo {author} {\bibfnamefont {M.}~\bibnamefont
  {Kamionkowski}}, \bibinfo {author} {\bibfnamefont {A.}~\bibnamefont
  {Kosowsky}}, \ and\ \bibinfo {author} {\bibfnamefont {D.~N.}\ \bibnamefont
  {Spergel}},\ }\href {\doibase 10.1103/PhysRevD.54.1332} {\bibfield  {journal}
  {\bibinfo  {journal} {Phys.Rev.}\ }\textbf {\bibinfo {volume} {D54}},\
  \bibinfo {pages} {1332} (\bibinfo {year} {1996})},\ \Eprint
  {http://arxiv.org/abs/astro-ph/9512139} {arXiv:astro-ph/9512139 [astro-ph]}
  \BibitemShut {NoStop}%
\bibitem [{\citenamefont {Fisher}(1935)}]{Fisher:1935e}%
  \BibitemOpen
  \bibfield  {author} {\bibinfo {author} {\bibfnamefont {R.~A.}\ \bibnamefont
  {Fisher}},\ }\href {http://www.jstor.org/stable/2342435} {\bibfield
  {journal} {\bibinfo  {journal} {Journal of the Royal Statistical Society}\
  }\textbf {\bibinfo {volume} {98}},\ \bibinfo {pages} {pp. 39} (\bibinfo
  {year} {1935})}\BibitemShut {NoStop}%
\bibitem [{\citenamefont {Ade}\ \emph {et~al.}(2015{\natexlab{b}})\citenamefont
  {Ade} \emph {et~al.}}]{Planck:2015xua}%
  \BibitemOpen
  \bibfield  {author} {\bibinfo {author} {\bibfnamefont {P.}~\bibnamefont
  {Ade}} \emph {et~al.} (\bibinfo {collaboration} {Planck Collaboration}),\
  }\href@noop {} {\  (\bibinfo {year} {2015}{\natexlab{b}})},\ \Eprint
  {http://arxiv.org/abs/1502.01589} {arXiv:1502.01589 [astro-ph.CO]}
  \BibitemShut {NoStop}%
\bibitem [{\citenamefont {Seo}\ and\ \citenamefont
  {Eisenstein}(2003)}]{Seo:2003pu}%
  \BibitemOpen
  \bibfield  {author} {\bibinfo {author} {\bibfnamefont {H.-J.}\ \bibnamefont
  {Seo}}\ and\ \bibinfo {author} {\bibfnamefont {D.~J.}\ \bibnamefont
  {Eisenstein}},\ }\href {\doibase 10.1086/379122} {\bibfield  {journal}
  {\bibinfo  {journal} {Astrophys.J.}\ }\textbf {\bibinfo {volume} {598}},\
  \bibinfo {pages} {720} (\bibinfo {year} {2003})},\ \Eprint
  {http://arxiv.org/abs/astro-ph/0307460} {arXiv:astro-ph/0307460 [astro-ph]}
  \BibitemShut {NoStop}%
\bibitem [{\citenamefont {Font-Ribera}\ \emph {et~al.}(2014)\citenamefont
  {Font-Ribera}, \citenamefont {McDonald}, \citenamefont {Mostek},
  \citenamefont {Reid}, \citenamefont {Seo} \emph
  {et~al.}}]{Font-Ribera:2013rwa}%
  \BibitemOpen
  \bibfield  {author} {\bibinfo {author} {\bibfnamefont {A.}~\bibnamefont
  {Font-Ribera}}, \bibinfo {author} {\bibfnamefont {P.}~\bibnamefont
  {McDonald}}, \bibinfo {author} {\bibfnamefont {N.}~\bibnamefont {Mostek}},
  \bibinfo {author} {\bibfnamefont {B.~A.}\ \bibnamefont {Reid}}, \bibinfo
  {author} {\bibfnamefont {H.-J.}\ \bibnamefont {Seo}},  \emph {et~al.},\
  }\href {\doibase 10.1088/1475-7516/2014/05/023} {\bibfield  {journal}
  {\bibinfo  {journal} {JCAP}\ }\textbf {\bibinfo {volume} {1405}},\ \bibinfo
  {pages} {023} (\bibinfo {year} {2014})},\ \Eprint
  {http://arxiv.org/abs/1308.4164} {arXiv:1308.4164 [astro-ph.CO]} \BibitemShut
  {NoStop}%
\bibitem [{\citenamefont {Abramo}\ and\ \citenamefont
  {Leonard}(2013)}]{Abramo:2013awa}%
  \BibitemOpen
  \bibfield  {author} {\bibinfo {author} {\bibfnamefont {L.~R.}\ \bibnamefont
  {Abramo}}\ and\ \bibinfo {author} {\bibfnamefont {K.~E.}\ \bibnamefont
  {Leonard}},\ }\href {\doibase 10.1093/mnras/stt465} {\bibfield  {journal}
  {\bibinfo  {journal} {Mon.Not.Roy.Astron.Soc.}\ }\textbf {\bibinfo {volume}
  {432}},\ \bibinfo {pages} {318} (\bibinfo {year} {2013})},\ \Eprint
  {http://arxiv.org/abs/1302.5444} {arXiv:1302.5444 [astro-ph.CO]} \BibitemShut
  {NoStop}%
\end{thebibliography}%

\end{document}